%% file: main.tex
\documentclass[conference,compsoc]{IEEEtran}
\IEEEoverridecommandlockouts

\usepackage{booktabs}
\usepackage{multirow}
\usepackage{tabularx}
\usepackage{array}
\usepackage{makecell}
\usepackage{colortbl}
\usepackage[table]{xcolor}

\usepackage{amsmath}
\usepackage{graphicx}
\usepackage{url}
\usepackage[protrusion=true,expansion=true]{microtype}

\usepackage{enumitem}
\usepackage{float}
\usepackage{placeins}
\usepackage{capt-of}

\usepackage{tikz}
\usepackage{fontawesome5}
\usetikzlibrary{
    shapes.geometric,
    arrows.meta,
    positioning,
    fit,
    backgrounds,
    calc,
    matrix
}

\usepackage[most]{tcolorbox}

\usepackage{rotating}

\newcolumntype{L}[1]{>{\raggedright\arraybackslash}p{#1}}

\definecolor{sysblue}{RGB}{35,35,35}
\definecolor{attackred}{RGB}{120,120,120}
\definecolor{corebrown}{RGB}{70,70,70}
\definecolor{actblue}{RGB}{50,50,50}
\definecolor{memblue}{RGB}{80,80,80}
\definecolor{uipbgreen}{RGB}{100,100,100}
\definecolor{toolpurple}{RGB}{60,60,60}
\definecolor{extorange}{RGB}{130,130,130}

\usetikzlibrary{
    shapes.geometric,
    arrows.meta,
    positioning,
    fit,
    backgrounds,
    calc,
    matrix
}

\tikzset{
  corebox/.style={
    rectangle,
    rounded corners,
    draw=corebrown!80!black,
    fill=corebrown!8,
    thick,
    minimum width=3.55cm,
    minimum height=1.32cm,
    inner sep=5pt,
    align=center
  },
  corebgbox/.style={
    rectangle,
    rounded corners,
    draw=corebrown!80!black,
    fill=corebrown!4,
    thick,
    inner sep=16pt
  },
  actbox/.style={
    rectangle,
    rounded corners,
    draw=actblue!80!black,
    fill=actblue!7,
    thick,
    minimum width=3.45cm,
    minimum height=1.22cm,
    inner sep=5pt,
    align=center
  },
  actbgbox/.style={
    rectangle,
    rounded corners,
    draw=actblue!80!black,
    fill=actblue!3,
    thick,
    inner sep=14pt
  },
  membox/.style={
    rectangle,
    rounded corners,
    draw=memblue!80!black,
    fill=memblue!7,
    thick,
    minimum width=3.1cm,
    minimum height=1.18cm,
    inner sep=5pt,
    align=center
  },
  membgbox/.style={
    rectangle,
    rounded corners,
    draw=memblue!80!black,
    fill=memblue!10,
    very thick,
    inner sep=16pt
  },
  uibox/.style={
    rectangle,
    rounded corners,
    draw=uipbgreen!80!black,
    fill=uipbgreen!8,
    thick,
    minimum height=1.08cm,
    inner sep=5pt,
    align=center,
    font=\bfseries
  },
  sysbox/.style={
    rectangle,
    rounded corners,
    draw=sysblue!80!black,
    fill=sysblue!10,
    fill opacity=0.35,
    text opacity=1,
    very thick,
    inner sep=18pt
  },
  toolbox/.style={
    rectangle,
    rounded corners,
    draw=toolpurple!90!black,
    fill=toolpurple!16,
    very thick,
    minimum width=2.65cm,
    inner sep=5pt,
    align=center
  },
  extbox/.style={
    rectangle,
    rounded corners,
    draw=extorange!80!black,
    fill=extorange!7,
    thick,
    minimum width=3.65cm,
    minimum height=1.22cm,
    inner sep=5pt,
    align=center
  },
  extbgbox/.style={
    rectangle,
    rounded corners,
    draw=extorange!55,
    fill=extorange!3,
    dashed,
    thick,
    inner sep=14pt
  },
  otheragent/.style={
    circle,
    draw=sysblue!80!black,
    fill=sysblue!8,
    thick,
    minimum size=1.15cm,
    align=center
  },
  attackbox/.style={
    rectangle,
    rounded corners,
    draw=attackred!85!black,
    fill=attackred!8,
    thick,
    inner sep=5pt,
    align=center,
    text width=3.7cm,
    minimum height=1.05cm
  },
  flow/.style={
    -{Stealth[scale=1.0]},
    draw=black!70,
    thick
  },
  flowright/.style={
    -{Stealth[scale=1.0]},
    draw=black!65,
    thick
  },
  internalflow/.style={
    -{Stealth[scale=1.0]},
    draw=gray!70,
    thick,
    dashed
  },
  attackarrow/.style={
    -{Stealth[scale=1.0]},
    draw=attackred!85!black,
    very thick,
    dashed
  }
}

\newtcolorbox{keyfindingbox}[1]{%
  enhanced,
  colback=white,
  colframe=black,
  colbacktitle=black!65,
  coltitle=white,
  fonttitle=\bfseries,
  title={#1},
  boxrule=1pt,
  titlerule=0pt,
  arc=1pt,
  outer arc=2pt,
  left=5pt,
  right=5pt,
  top=5pt,
  bottom=5pt
}

\begin{document}

\setlength{\parskip}{0pt}

\title{Agents of Chaos: A Systematization of Knowledge (SoK) on Malicious Attacks on Agentic AI Systems}


\author{
\IEEEauthorblockN{
Thanh M. Truong-Le\IEEEauthorrefmark{1},
Amber Bennoui\IEEEauthorrefmark{2},
Hoang-Ly Nguyen\IEEEauthorrefmark{3},
Gia-Thang Ho\IEEEauthorrefmark{4},\\
Thai-Minh Truong\IEEEauthorrefmark{4},
Duc-Ly Vu\IEEEauthorrefmark{5}\textsuperscript{1}%
\thanks{\textsuperscript{1}\,Corresponding author:
Duc-Ly Vu (e-mail: lyvd@tdmu.edu.vn).}
}
\IEEEauthorblockA{
\IEEEauthorrefmark{1}University of Information Technology, Vietnam\\
thanhtlm.19@grad.uit.edu.vn\\
\IEEEauthorrefmark{2}Jiffy Labs, USA\\
amber@jiffylabs.ai\\
\IEEEauthorrefmark{3}Eastern International University, Vietnam\\
ly.nguyen@eiu.edu.vn\\
\IEEEauthorrefmark{4}Ho Chi Minh City University of Technology (HCMUT),
Vietnam National University, Vietnam\\
thang.ho261104@hcmut.edu.vn; thaiminh@hcmut.edu.vn\\
\IEEEauthorrefmark{5}Thu Dau Mot University, Vietnam\\
lyvd@tdmu.edu.vn
}
}
\maketitle

\begin{abstract}
Agentic AI systems extend large language models with tools, memory, retrieval, browser or computer-use interfaces, code execution, external services, and multi-agent coordination. Although these capabilities enable agents to act beyond text generation and perform tasks, they also create attack paths through which adversaries can exploit untrusted content, deceptive metadata, poisoned state, compromised skills, mutable dependencies, or malicious code. Such attacks can manipulate planning and tool use, corrupt memory, abuse code-execution environments, and trigger harmful downstream effects.

In this paper, we present a systematic review of malicious attacks targeting agentic AI systems. We screened 186 candidate sources and retained 98. Following independent coding and resolver adjudication, we identified 284 atomic attacks, 452 atomic safeguards, and 660 explicit attack--safeguard relationships. We organize attacks into eight causal-mechanism classes while retaining insertion points, activation conditions, abused capabilities, objectives, impacts, persistence, and evidence as orthogonal dimensions.

Our analysis shows that tool, MCP, and action-layer attacks account for the largest share of atomic attacks, while the prompt-injection and action-hijacking class has the broadest paper coverage. We identify distinct class-specific activation patterns and separation between insertion and later activation. Safeguard coverage remains uneven: only 278 of 660 relationships are directly assessed, and strong coverage is uncommon. These findings support lifecycle-aware threat modeling, agent evaluation, tool governance, and defense-in-depth security for agentic AI systems.
\end{abstract}



\input{sections/01-introduction}

\input{sections/02-background-scope}
\input{sections/03-methodology}
\input{sections/04-taxonomy}
\input{sections/05-incidents-observations}
\input{sections/06-safeguards}
\input{sections/07-research-gaps}
\input{sections/08-related-work}
\input{sections/09-threats-ethics}
\input{sections/10-conclusion}
\clearpage

%


\bibliographystyle{IEEEtran}
\bibliography{references}

%


\appendices
\input{sections/appendix.tex}

\end{document}

%% file: sections/01-introduction.tex
\section{Introduction}
\label{sec:introduction}

Modern agentic AI systems are no longer passive text generators. Unlike standalone language models, which primarily receive an input and produce an output, agentic AI systems can observe external content, reason over intermediate state, choose actions, and cause side effects outside the model~\cite{wang2024survey,chhabra2026agenticsecurity,
dehghantanha2026attacksurface,kim2026attackdefenselandscape}. To do so, agentic AI systems extend large language models with tools, memory, retrieval, browser and computer-use interfaces, code execution, external services, and multi-agent coordination~\cite{dehghantanha2026attacksurface,
kim2026attackdefenselandscape,wang2024survey}. 
These capabilities expand the security boundary beyond
the model to workflows connecting instructions, data,
tools, memory, artifacts, permissions, and external
services~\cite{chhabra2026agenticsecurity,
kim2026attackdefenselandscape,
dehghantanha2026attacksurface}.

This expanded perimeter introduces attack surfaces that differ substantially from those of standalone models, partly because input channels can serve as injection vectors~\cite{greshake2023notwhat,chhabra2026agenticsecurity,dehghantanha2026attacksurface}. For example, malicious instructions embedded in webpages, emails, documents, tool responses, retrieved passages, memory entries, or skill files can influence how an agent plans and acts~\cite{greshake2023notwhat,
liu2023promptinjection,schmotz2026skillinject,liu2026maliciousskills}. Deceptive tool descriptions or malicious Model Context Protocol (MCP) servers can bias tool selection and parameter generation~\cite{wang2026mcptox,dehghantanha2026attacksurface,
kim2026attackdefenselandscape}. Poisoned memory entries or retrieval content can persist across sessions and influence future decisions~\cite{chen2024agentpoison,zou2025poisonedrag}. Malicious skills, dependencies, or installers can cause agents to access secrets, execute commands, modify files, or communicate with external endpoints~\cite{liu2026maliciousskills,schmotz2026skillinject,
ji2026cloak,ladisa2023sok,zhao2024models}. In multi-agent settings, compromised instructions or outputs can propagate through communication channels and collaborative workflows~\cite{lee2024promptinfection,arif2026conjunctive}.

As in software supply chains, reusable artifacts such as agent skills, plugins, MCP servers, prompts, workflows, datasets, model components, configuration files, and dependencies can deliver malicious content across trust boundaries before integration into agentic systems~\cite{ladisa2023sok,zhao2024models,liu2026maliciousskills,schmotz2026skillinject}. However, the central security challenge
extends beyond artifact provenance to how
untrusted content and compromised components
influence agent reasoning, memory, tool use,
permissions, and downstream effects.

Research on these attacks remains fragmented.
Recent surveys and SoKs cover agent design,
attack surfaces, threat models, defenses,
evaluation, benchmarks, governance, and
deployment~\cite{chhabra2026agenticsecurity,
dehghantanha2026attacksurface,kim2026attackdefenselandscape}. However,
they do not provide a focused, cross-mechanism systematization of
malicious attack vectors and corresponding defenses. 

This gap motivates
the following three research questions in this paper:

\begin{itemize}
    \item \textbf{RQ1:} What is a comprehensive list of malicious attack vectors targeting agentic AI?

    \item \textbf{RQ2:} What patterns emerge across documented malicious attack vectors?

    \item \textbf{RQ3:} Which safeguards address each attack class, and where do significant coverage gaps remain?
\end{itemize}
To answer these questions, we conducted a systematic review of
peer-reviewed publications, preprints, platform advisories, repository
disclosures, incident reports, and other technically attributable grey
literature. Our final analytical corpus comprises 98 sources. Following
independent coding and resolver adjudication, we identified 284 atomic
attacks, 452 atomic safeguards, and 660 explicit source-supported
attack--safeguard relationships.

Our primary analytical unit is the atomic attack, not a source
quotation or broad thematic vector. We distinguish attacks when
their causal mechanism or attack surface differs materially, or
when differences in access, delivery, insertion, activation, or
an indispensable target define distinct attack processes.
Each attack receives one primary causal class, while target,
delivery, insertion, activation, capability, objective, impact,
persistence, and evidence remain orthogonal dimensions.

For RQ2, we analyze relationships among insertion point, activation
condition, abused capability, objective, impact, and persistence. For
RQ3, we treat safeguards as atomic controls and separately record their
control type, evaluation status, origin, and enforcement target. We
retain only explicit source-supported attack--safeguard relationships
and distinguish whether each relationship was directly assessed and,
when evaluated, whether its coverage was strong, partial, or limited.

The resulting taxonomy and safeguard mapping
offer a mechanism-based view of agentic-AI
attacks and reveal gaps in defensive evidence.

%% file: sections/02-background-scope.tex
\section{Background}
\label{sec:background}

\subsection{Agentic AI Systems and Risks}
\label{sec:agentic_components}

We use the term \emph{agentic AI system} to refer to an AI-enabled
system that pursues goals through multiple steps by interpreting
instructions, maintaining state, observing its environment, selecting
actions, and using action outcomes to inform subsequent decisions. Unlike
a standalone language model, an agentic system may retrieve external
information, invoke tools or APIs, execute code, interact with browser or
computer-use interfaces, and coordinate with other agents, thereby causing
side effects beyond the model itself~\cite{wang2024survey,
chhabra2026agenticsecurity,dehghantanha2026attacksurface,
kim2026attackdefenselandscape}. This shift expands the security boundary beyond the model to the
surrounding workflow. Prompt injection and context contamination exploit
instruction--data confusion within the model context~\cite{greshake2023notwhat,
liu2023promptinjection}. Tool and MCP attacks manipulate descriptions,
schemas, registries, or runtime behavior to influence tool selection and
invocation~\cite{wang2026mcptox}. Memory and retrieval attacks can preserve
malicious influence across sessions or activate when poisoned content is
later recalled~\cite{chen2024agentpoison,zou2025poisonedrag}. Browser,
computer-use, and coding-agent attacks can translate model outputs into
commands, file edits, package installations, or external
requests~\cite{isbarov2026agentproxy,ji2026cloak,krishna2025phantoms}.
Multi-agent attacks allow malicious instructions or outputs to propagate
through communication and delegation
channels~\cite{lee2024promptinfection,arif2026conjunctive}.

Prompts, skills, workflows, MCP servers, model
components, configurations, and dependencies
can carry malicious content across trust
boundaries~\cite{liu2026maliciousskills,
schmotz2026skillinject,wang2026mcptox,ladisa2023sok,zhao2024models}.
We treat these artifacts as part of the broader
agentic-AI attack surface, not its sole focus.
\subsection{Attack Lifecycle: From Insertion to Activation}
\label{sec:attack_lifecycle}
A recurring property of agentic-AI attacks is that insertion and
activation may be separated in time and occur across different
components~\cite{chen2024agentpoison,zou2025poisonedrag,
wang2024badagent,yang2024watch,wang2026mcptox,
liu2026maliciousskills,schmotz2026skillinject,ji2026cloak}. We model these
attacks using a six-stage lifecycle. First, during \emph{insertion}, an
attacker introduces malicious or misleading content into a webpage,
prompt, tool description, memory entry, retrieval corpus, skill file,
dependency, or other artifact. Second, during \emph{exposure and
selection}, the agent encounters this content through retrieval, browsing,
registry search, tool discovery, installation, or inter-agent
communication. Third, during \emph{interpretation and planning}, the agent
may process the untrusted content as an instruction, evidence, or
authority. Fourth, during \emph{activation and action}, the malicious
content triggers side effects such as tool invocation, command execution,
file modification, dependency installation, or data
exfiltration~\cite{greshake2023notwhat,liu2023promptinjection,
isbarov2026agentproxy,ji2026cloak}. Fifth, \emph{persistence and
propagation} may occur through memory, dependencies, backdoors, installed
components, or multi-agent communication~\cite{chen2024agentpoison,
wang2024badagent,yang2024watch,lee2024promptinfection,
arif2026conjunctive}. Finally, \emph{detection, containment, and recovery}
involve safeguards such as prompt guards, tool validation, runtime
monitoring, sandboxing, least privilege, memory cleanup, provenance checks,
and revocation~\cite{chhabra2026agenticsecurity,
dehghantanha2026attacksurface,kim2026attackdefenselandscape,
bennet2024aibom}.

\subsection{Trust Boundaries and Stakeholders}
\label{sec:trust_boundaries}
Trust boundaries arise whenever content, authority, or state moves
between entities with different trust assumptions or privilege levels. In
agentic AI systems, key boundaries include external-content-to-context,
retrieval-to-agent, memory-to-session, agent-to-tool, agent-to-runtime,
agent-to-agent, and registry-to-developer
transitions~\cite{chhabra2026agenticsecurity,
dehghantanha2026attacksurface,kim2026attackdefenselandscape,
ladisa2023sok}. Stakeholders across these boundaries include developers,
tool and MCP providers, skill authors, curators, maintainers, registry
operators, administrators, downstream users, and affected data
subjects~\cite{bennet2024aibom,liu2026maliciousskills,
schmotz2026skillinject,zhao2024models}.



%% file: sections/03-methodology.tex
\section{Methodology}
\label{sec:methodology}

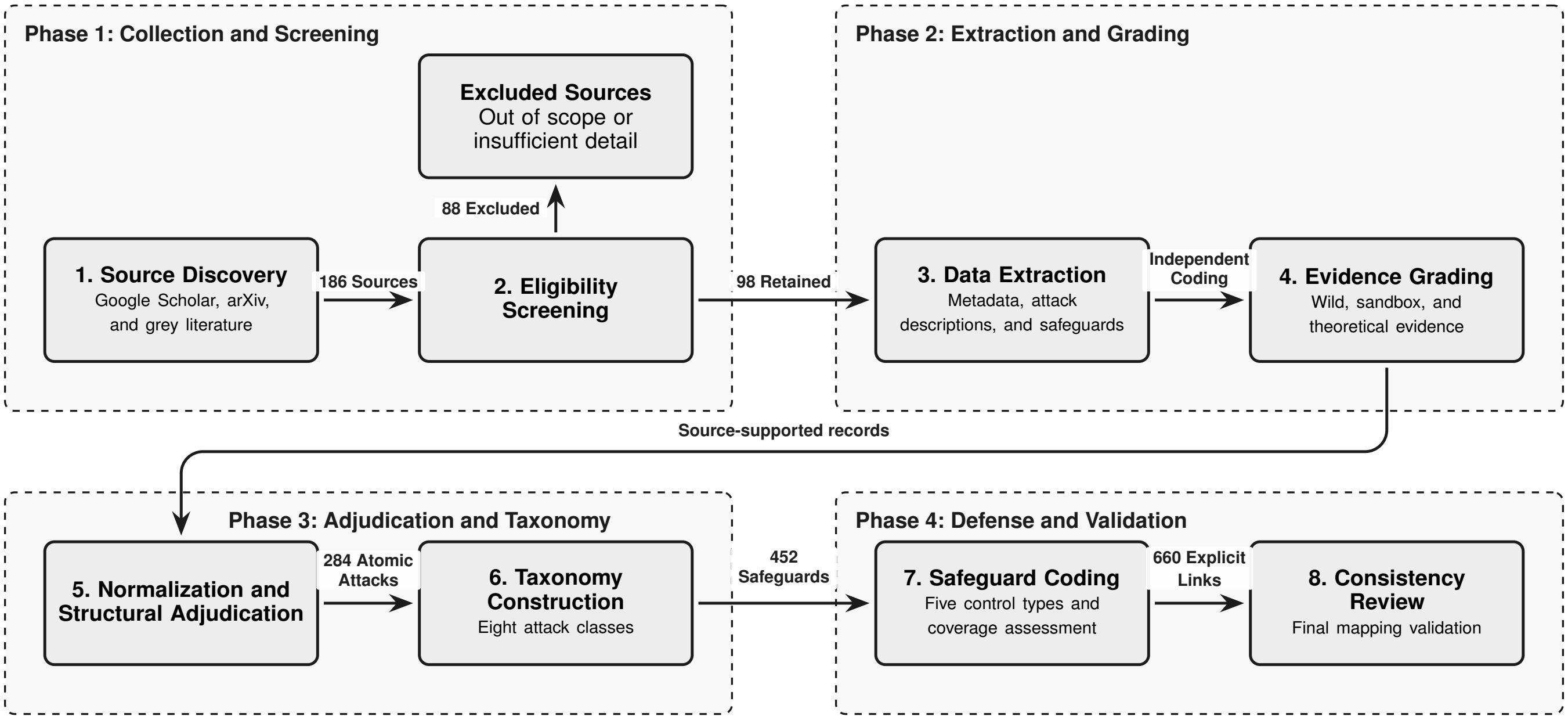
\begin{figure*}[t]
\centering
\resizebox{\textwidth}{!}{%
\begin{tikzpicture}[
    every node/.append style={font=\sffamily},
    workflowbox/.style={
        rectangle,
        rounded corners,
        draw=sysblue!80!black,
        fill=sysblue!8,
        very thick,
        minimum width=4.0cm,
        minimum height=1.8cm,
        text width=3.7cm,
        align=center,
        font=\sffamily\small
    },
    bluebox/.style={workflowbox},
    redbox/.style={workflowbox},
    phase3box/.style={workflowbox},
    phase4box/.style={workflowbox},
    excludebox/.style={workflowbox},
    bluearrow/.style={
        -{Stealth[scale=1.2]},
        very thick,
        draw=sysblue!80!black,
        shorten <=2pt,
        shorten >=2pt
    },
    redarrow/.style={
        -{Stealth[scale=1.2]},
        very thick,
        draw=sysblue!80!black,
        shorten <=2pt,
        shorten >=2pt
    },
    bluelabel/.style={
        font=\sffamily\scriptsize\bfseries,
        text=sysblue!90!black,
        fill=white,
        fill opacity=0.96,
        text opacity=1,
        rounded corners=1pt,
        inner xsep=3pt,
        inner ysep=1.5pt,
        align=center
    },
    redlabel/.style={
        font=\sffamily\scriptsize\bfseries,
        text=sysblue!90!black,
        fill=white,
        fill opacity=0.96,
        text opacity=1,
        rounded corners=1pt,
        inner xsep=3pt,
        inner ysep=1.5pt,
        align=center
    },
    phasebox/.style={
        rectangle,
        draw=sysblue!70!black,
        dashed,
        inner xsep=16pt,
        inner ysep=20pt,
        rounded corners,
        fill=sysblue!3,
        thick
    },
    phaseblue/.style={phasebox},
    phasered/.style={phasebox},
    phasetitle/.style={
        font=\sffamily\small\bfseries
    }
]

\node[bluebox] (S1) {
    \textbf{1. Source Discovery}
    \\
    {\mdseries\scriptsize Google Scholar, arXiv, and grey literature}
};

\node[bluebox, right=1.45cm of S1] (S2) {
    \textbf{2. Eligibility}
    \\
    \textbf{Screening}
};

\node[excludebox, above=0.85cm of S2] (Excluded) {
    \textbf{Excluded Sources}
    \\
    {\mdseries Out of scope or insufficient detail}
};

\node[redbox, right=2.65cm of S2] (S3) {
    \textbf{3. Data Extraction}
    \\
    {\mdseries\scriptsize Metadata, attack descriptions, and safeguards}
};

\node[redbox, right=1.45cm of S3] (S4) {
    \textbf{4. Evidence Grading}
    \\
    {\mdseries\scriptsize Wild, sandbox, and theoretical evidence}
};

\node[phase3box, below=2.6cm of S1] (S5) {
    \textbf{5. Normalization and}
    \\
    \textbf{Structural Adjudication}
};

\node[phase3box] (S6) at (S2 |- S5) {
    \textbf{6. Taxonomy}
    \\
    \textbf{Construction}
    \\
    {\mdseries\scriptsize Eight attack classes}
};

\node[phase4box] (S7) at (S3 |- S5) {
    \textbf{7. Safeguard Coding}
    \\
    {\mdseries\scriptsize Five control types and coverage assessment}
};

\node[phase4box] (S8) at (S4 |- S5) {
    \textbf{8. Consistency}
    \\
    \textbf{Review}
    \\
    {\mdseries\scriptsize Final mapping validation}
};

\coordinate (P2TopLeft) at (S3 |- Excluded.north);
\coordinate (P2TopRight) at (S4 |- Excluded.north);

\begin{scope}[on background layer]

    \node[phaseblue, fit=(S1) (S2) (Excluded)] (P1) {};
    \node[
        phasetitle,
        text=sysblue!80!black,
        anchor=north west,
        xshift=5pt,
        yshift=-5pt
    ] at (P1.north west)
    {Phase 1: Collection and Screening};

    \node[phasered, fit=(S3) (S4) (P2TopLeft) (P2TopRight)] (P2) {};
    \node[
        phasetitle,
        text=sysblue!80!black,
        anchor=north west,
        xshift=5pt,
        yshift=-5pt
    ] at (P2.north west)
    {Phase 2: Extraction and Grading};
    \node[phaseblue, fit=(S5) (S6)] (P3) {};
    
    \node[
        phasetitle,
        text=sysblue!80!black,
        anchor=north west,
        xshift=90pt,
        yshift=-5pt
    ] at (P3.north west)
    {Phase 3: Adjudication and Taxonomy};

    \node[phasered, fit=(S7) (S8)] (P4) {};
    \node[
        phasetitle,
        text=sysblue!80!black,
        anchor=north west,
        xshift=5pt,
        yshift=-5pt
    ] at (P4.north west)
    {Phase 4: Defense and Validation};

\end{scope}

\draw[bluearrow]
    (S1.east) --
    node[bluelabel, midway, above=3pt] {186 Sources}
    (S2.west);

\draw[redarrow]
    (S2.north) --
    node[redlabel, midway, left=3pt] {88 Excluded}
    (Excluded.south);

\draw[redarrow]
    (S2.east) --
    node[redlabel, midway, above=3pt] {98 Retained}
    (S3.west);
\draw[redarrow]
    (S3.east) --
    node[redlabel, midway, above=3pt, align=center]
    {Independent\\Coding}
    (S4.west);

\coordinate (ReturnWest) at ($(S5.north)+(0,1.3cm)$);
\coordinate (ReturnEast) at (S4.south |- ReturnWest);
\draw[bluearrow, rounded corners=9pt]
    (S4.south) --
    (ReturnEast) --
    node[bluelabel, midway, above=3pt] {Source-supported records}
    (ReturnWest) --
    (S5.north);
\draw[bluearrow]
    (S5.east) --
    node[bluelabel, midway, above=5pt]
    {\shortstack{284 Atomic\\Attacks}}
    (S6.west);
\draw[redarrow]
    (S6.east) --
    node[redlabel, midway, above=5pt]
    {\shortstack{452\\Safeguards}}
    (S7.west);
\draw[redarrow]
    (S7.east) --
    node[redlabel, midway, above=5pt]
    {\shortstack{660 Explicit\\Links}}
    (S8.west);

\end{tikzpicture}%
}

\caption{Workflow for source collection, evidence extraction, structural
adjudication, taxonomy construction, safeguard coding, and validation.}
\label{fig:workflow}
\end{figure*}

Our methodology comprises eight sequential stages, as illustrated in
Figure~\ref{fig:workflow}: source discovery, eligibility screening,
data extraction, evidence grading, normalization and structural
adjudication, taxonomy construction, safeguard coding, and consistency
review.

We considered peer-reviewed literature and technically attributable grey
literature to capture both archival studies and early disclosures of
agentic-AI attacks. A source was eligible if it (1) described an agentic
AI system, component, workflow, artifact, or related security mechanism;
(2) reported an attack that affected, or was designed to affect, agentic
behavior or downstream actions; and (3) provided sufficient technical
detail to identify a concrete attack path, security-relevant effect, or
safeguard.

We included attacks that targeted an agentic workflow and produced, or
were designed to produce, a security-relevant effect beyond an
undesirable text response. We excluded ordinary model errors,
hallucinations, bias, general performance degradation, direct adversarial
examples without downstream agentic action, and speculative discussions
that did not specify a concrete attack path or security impact.

The literature collection was bounded by a
July 2026 cutoff. Sources published or
disclosed after this cutoff were not included
in the analytical corpus.

Google Scholar and arXiv served as the primary discovery interfaces. We
supplemented these searches with backward and forward snowballing,
targeted venue searches, and technically attributable grey literature,
including platform disclosures, repository reports, security advisories,
vulnerability records, and incident reports. Because terminology varies
across prompt injection, tool and MCP attacks, memory and RAG poisoning,
multi-agent security, backdoors, malicious skills, and AI supply-chain
security, we used multiple query families rather than a single search
string. Table~\ref{tab:search_queries} reports the corresponding
candidate and retained source counts.

\begin{table*}[t]
\centering
\caption{\textsc{Our search queries and screening outcomes.}}
\label{tab:search_queries}

\small
\renewcommand{\arraystretch}{1.04}
\setlength{\tabcolsep}{5pt}

\begin{tabularx}{\textwidth}{
    >{\raggedright\arraybackslash}p{0.205\textwidth}
    >{\raggedright\arraybackslash}X
    >{\centering\arraybackslash}p{0.075\textwidth}
    >{\centering\arraybackslash}p{0.060\textwidth}
}
\toprule
\textbf{Query Category}
&
\textbf{Search Query}
&
\textbf{Candidates}
&
\textbf{Selected}
\\
\midrule

\rowcolor{gray!10}
Broad agentic attack search
&
(``agentic AI'' OR ``LLM agent*'' OR ``tool-using agent*''
OR ``multi-agent LLM*'' OR ``computer-use agent*'')
AND
(attack* OR malicious OR poison* OR backdoor*
OR hijack* OR inject* OR security OR safeguard
OR defense OR governance)
&
71
&
34
\\

Indirect prompt injection, MCP, and tool security
&
(``LLM agent*'' OR ``tool-using agent*''
OR ``Model Context Protocol'' OR MCP)
AND
(``indirect prompt injection'' OR ``tool poisoning''
OR ``tool hijacking'' OR malicious);
supplemented with
``indirect prompt injection'' AND (LLM OR agent) AND defense
&
22
&
18
\\

\rowcolor{gray!10}
Memory and agentic RAG poisoning
&
(``LLM agent*'' OR ``agentic RAG'')
AND
(``memory poisoning'' OR ``memory injection''
OR ``knowledge-base poisoning'' OR ``RAG poisoning''
OR ``persistent memory compromise'')
&
19
&
12
\\

Multi-agent attacks and propagation
&
(``multi-agent LLM*'' OR ``agent-to-agent''
OR ``agent collaboration'' OR ``agent communication'')
AND
(``malicious agent'' OR corruption OR poisoning
OR backdoor* OR propagation OR collusion)
&
8
&
4
\\

\rowcolor{gray!10}
Computer-use, browser, web, and coding agents
&
(``computer-use agent*'' OR ``browser agent*''
OR ``web agent*'' OR ``coding agent*''
OR ``agentic coding assistant*'')
AND
(attack* OR injection OR hijack* OR exfiltrat*
OR ``command execution'')
&
8
&
8
\\

Privacy, exfiltration, and agent backdoors
&
(``LLM agent*'' OR ``tool-using agent*''
OR ``autonomous agent*'')
AND
(``data exfiltration'' OR ``privacy leakage''
OR credential* OR backdoor* OR ``persistent compromise'')
&
8
&
3
\\

\rowcolor{gray!10}
Agent skills, AI supply chains, and artifact security
&
(``agent skill'' OR ``skill file'' OR ``SKILL.md''
OR plugin OR package OR dependency OR repository
OR registry OR ``model hub'' OR ``AI supply chain'')
AND
(malicious OR attack OR poisoning OR backdoor
OR injection OR ``scanner evasion'' OR provenance
OR ``supply chain'')
&
50
&
19
\\

\addlinespace[2pt]
\multicolumn{2}{r}{\textbf{Total}}
&
\textbf{186}
&
\textbf{98}
\\
\bottomrule
\end{tabularx}

\vspace{2pt}
\begin{minipage}{\textwidth}
\footnotesize
A collected record could match multiple keyword families. To make
the row totals additive, each record was assigned to the family that
best represented its primary topic. The counts therefore describe
the locally collected candidate corpus rather than the total number
of results displayed by Google Scholar or arXiv. Search terms related
to safeguards, defenses, governance, provenance, sandboxing, least
privilege, monitoring, and revocation were incorporated into the
broad agentic-security query family.
\end{minipage}

\end{table*}

\subsection{Normalization, Atomicity, and Deduplication}

Different sources often use different terms for similar mechanisms, while
the same study may describe multiple payloads, experimental settings,
objectives, or downstream effects. We therefore distinguish source
evidence from analytical records. Quotations, highlighted passages,
benchmark results, and source descriptions provide evidence for coding
but are not themselves counting units.

Our primary attack unit is an \emph{atomic attack record}. We split two
observations into distinct attacks only when their causal mechanism or
attack surface materially differs, or when differences in attacker
access, delivery, insertion point, activation condition, or an
indispensable target define a different attack process. Differences only
in terminology, prompt wording, victim model, experimental
configuration, objective, or downstream impact do not by themselves
justify separate attack records.

For example, malicious instructions delivered through webpages, emails,
documents, or tool outputs may represent the same prompt-injection
mechanism when they exploit the same instruction--data boundary.
Conversely, two attacks delivered through the same artifact are treated
as distinct when one depends on semantic instruction following and the
other depends on executable code, package installation, or malicious
runtime behavior.

We apply the same atomicity principle to safeguards. A safeguard is
counted separately when the source presents it as a distinct control
mechanism or independently evaluates it as a defense baseline. Internal
implementation stages are not automatically counted as separate
safeguards unless they enforce causally distinct security boundaries.

After structural adjudication, the final analytical dataset contains
284 atomic attacks and 452 atomic safeguards.

\subsection{Taxonomy Construction}

We constructed the taxonomy iteratively through bottom-up mechanism
coding and top-down consistency review. During bottom-up analysis, we
extracted the causal process through which each attack influences an
agentic system. During top-down analysis, we grouped related mechanisms
into eight primary attack classes.

Each atomic attack receives exactly one primary class based
on its decisive causal mechanism, while carriers, target
components, abused capabilities, objectives, and impacts
remain orthogonal dimensions. Primary assignments are
mutually exclusive, although mechanisms may overlap
within compound incidents.

The final taxonomy contains eight classes:
(1) prompt injection and action hijacking;
(2) tool, MCP, and action-layer attacks;
(3) memory and RAG poisoning;
(4) multi-agent attacks;
(5) backdoors and trigger-conditioned compromise;
(6) computer-use and coding-agent attacks;
(7) agent skills and supply-chain attacks; and
(8) goal, task, and planning manipulation.

To distinguish overlapping mechanisms, C1
captures instruction--data boundary violations,
whereas C2 applies when manipulation of tool or
MCP interfaces is the decisive mechanism.
C3 is assigned when poisoning persistent memory
or retrieval state is indispensable to the attack,
rather than merely serving as a delivery channel
for instructions. C8 captures manipulation of
goal selection, task state, or planning logic,
rather than ordinary instruction-following
hijacking under C1.

Similarly, C4 requires agent-to-agent
interaction or propagation to be central to
the compromise, rather than merely occurring in a multi-agent setting. C5 applies when an implanted
backdoor is activated by a trigger, even if
the resulting behavior involves tool use;
C2 applies when the tool or MCP interface
itself is manipulated. C6 captures attacks
whose decisive mechanism exploits
computer-use or coding-agent execution,
whereas C7 applies when compromise through
a reusable skill, package, dependency, or
other supply-chain artifact is essential
to the attack.

For compound incidents, structural adjudication
first determines whether the evidence supports
one causal attack process or multiple distinct
atomic attacks under Section~3.1. Crossing
multiple lifecycle stages or producing several
impacts does not alone justify splitting.
Each resulting atomic record receives one
primary causal class.

For the cross-cutting analysis in Section~5, we additionally code target
component, attacker access, delivery channel, insertion point,
activation condition, abused capability, attacker objective, downstream
impact, persistence or propagation, evidence context, evidence level,
and evidence type.
The complete set of attack, evidence, safeguard, and relationship-level
coding dimensions is summarized in the final codebook in
Section~\ref{sec:final_codebook}.

\subsection{Safeguard Coding and Coverage Assessment}
\label{sec:safeguard_coding}

We extracted safeguards from attack studies, empirical evaluations, system designs, benchmark papers, governance proposals, and dedicated defense papers, and classified them into five control types. First, \textit{preventive controls} seek to block attack insertion or unsafe selection through provenance checks, tool allowlists, input validation, artifact signing, and permission design. Second, \textit{detective controls} identify suspicious content or behavior through prompt-injection detectors, scanner-based review, anomaly detection, runtime tracing, and tool-call inspection. Third, \textit{containment controls} limit the effects of successful attacks through sandboxing, least privilege, network restrictions, scoped credentials, and human approval for sensitive actions. Fourth, \textit{corrective controls} remove or repair compromised state through memory cleanup, dependency rollback, poisoned-corpus removal, and agent replacement. Finally, \textit{recovery controls} support post-incident response through logging, revocation, downstream notification, audit trails, and incident response procedures.

For RQ3, safeguards are analyzed as atomic control records. In addition
to control type, we code each safeguard's status, origin, and direct
enforcement target.

We evaluate coverage at the level of explicit attack--safeguard
relationships rather than assuming that a safeguard applies uniformly
to every attack in the same class. A relationship is retained only when
the source explicitly links the safeguard to the attack mechanism or
when a direct defense experiment establishes that relationship.

For each explicit relationship, we record whether the safeguard was
directly assessed against the corresponding attack. Directly assessed
relationships are classified as \emph{strong} when the safeguard
directly addresses the core mechanism under the evaluated assumptions
and leaves little residual attack success; \emph{partial} when it
meaningfully reduces attack success or impact but leaves important
bypasses, uncovered stages, or restrictive assumptions; and
\emph{limited} when protection is weak, indirect, brittle, or highly
assumption-dependent. Explicit relationships without attack-specific
evaluation are coded as \emph{not evaluated} rather than receiving a
speculative effectiveness rating.

The final dataset contains 660 explicit source-supported
attack--safeguard relationships. The resulting attack, evidence, safeguard, and relationship-level
coding dimensions are consolidated in the final codebook described
next.
\subsection{Final Codebook}
\label{sec:final_codebook}
To support consistent analysis across the reviewed literature, we
developed a final codebook that separates the primary causal mechanism
from orthogonal attack, evidence, safeguard, and coverage dimensions.
The codebook operates at three analytical levels: atomic attacks, atomic
safeguards, and explicit attack--safeguard relationships.
Table~\ref{tab:codebook_groups} summarizes the six code groups used in
the analysis. Several dimensions are multi-valued because a single
atomic attack may involve multiple targets, capabilities, objectives, or
impacts. In contrast, each atomic attack receives exactly one primary
attack class based on its causal mechanism. The complete codebook,
including dimension definitions and representative coded examples, is
provided in Appendix~\ref{app:full-codebook}.

\begin{table}[t]
\centering
\caption{\textsc{Summary of the final code groups used in the study.}}
\label{tab:codebook_groups}
\small
\setlength{\tabcolsep}{4pt}
\renewcommand{\arraystretch}{1.06}

\begin{tabularx}{\columnwidth}{L{2.1cm}>{\raggedright\arraybackslash}X}
\toprule
\rowcolor{white}
\textbf{Code group} &
\textbf{Dimensions} \\
\midrule

\rowcolor{gray!10}
Attack classification
&
Primary class; causal mechanism \\

\rowcolor{white}
Attack path
&
Target component; attacker access; delivery channel; insertion point;
activation condition \\

\rowcolor{gray!10}
Agent behavior
&
Abused capability; attacker objective; downstream impact;
persistence / propagation \\

\rowcolor{white}
Evidence
&
Context; evidence level; evidence type \\

\rowcolor{gray!10}
Safeguard
&
Control type; status; origin; enforcement target \\

\rowcolor{white}
Coverage
&
Direct assessment; coverage level \\

\bottomrule
\end{tabularx}
\end{table}

\subsection{Inter-Coder Reliability and Adjudication}
Inter-coder reliability was computed from independent
pre-adjudication outputs using eligible aligned records
with valid paired assignments. The two coders identified
280 and 251 attack records, respectively. Among 367
unique pre-adjudication attack identifiers, 164 were
shared, of which 159 had complete primary-class
assignments. Safeguard coding produced 412 and 218
records, with 172 shared identifiers.

Unmatched records were excluded from field-level
agreement calculations and addressed during subsequent
structural adjudication. Thus, the reported agreement
reflects coding consistency on aligned records rather
than agreement on record discovery or atomic split/merge
decisions.

Table~\ref{tab:intercoder_reliability} summarizes pre-adjudication agreement on the main coding fields. Primary attack-class assignment achieved 78.62\% agreement ($\kappa = 0.75$) on independently coded, aligned records, supporting the consistency of the primary taxonomy within this comparison, whereas agreement was lower for some other dimensions, particularly safeguard enforcement targets (40.91\%, $\kappa = 0.37$), indicating greater assignment uncertainty. Resolver adjudication subsequently established a consistent final dataset, but these pre-adjudication results do not imply independent reproducibility across all coding dimensions.



\begin{table}[t]
\centering
\caption{\textsc{Pre-adjudication inter-coder agreement on eligible aligned records.}}
\label{tab:intercoder_reliability}
\small
\setlength{\tabcolsep}{2.2pt}
\renewcommand{\arraystretch}{1.05}

\begin{tabularx}{\columnwidth}{>{\raggedright\arraybackslash}X c c c}
\toprule
\rowcolor{white}
\textbf{Coding field} &
\textbf{Aligned pairs} &
\textbf{Agreement} &
\textbf{$\kappa$} \\
\midrule

\rowcolor{gray!10}
Primary attack class
& 159 & 78.62\% & 0.75 \\

\rowcolor{white}
Paper role
& 98 & 67.35\% & 0.26 \\

\rowcolor{gray!10}
Attack-evidence presence
& 98 & 88.78\% & 0.28 \\

\rowcolor{white}
Safeguard-evidence presence
& 98 & 82.65\% & 0.24 \\

\rowcolor{gray!10}
SoK framing
& 98 & 95.92\% & 0.48 \\

\rowcolor{white}
Safeguard control type
& 165 & 69.09\% & 0.41 \\

\rowcolor{gray!10}
Safeguard status
& 164 & 72.56\% & 0.55 \\

\rowcolor{white}
Safeguard origin
& 163 & 69.94\% & 0.51 \\

\rowcolor{gray!10}
Safeguard enforcement target
& 154 & 40.91\% & 0.37 \\

\bottomrule
\end{tabularx}
\end{table}













The empirical-component field showed 98.98\% raw agreement but
$\kappa\approx0$ because the category was nearly constant across the
corpus; this prevalence-sensitive result is therefore reported
separately rather than included in the table.

Resolver adjudication followed a fixed order:
record existence and atomic split/merge decisions;
primary attack class; attack dimensions;
evidence dimensions; safeguard structure and
fields; explicit attack--safeguard relationships;
and final quality assurance. During structural adjudication, candidate records
were reviewed against the source evidence and
frozen atomicity rules. Retained records were
assigned canonical identifiers, with pre-adjudication
identifiers mapped to their corresponding canonical
records. Unmatched and overlapping candidates were
resolved before field-level reconciliation.

Disagreements were resolved against the source
documents and frozen coding rules rather than
by treating either coder as ground truth. Appendix~\ref{app:adjudication-example} illustrates a documented
pre-adjudication disagreement and its
source-grounded resolution.

\subsection{Consistency Review and Reproducibility}

After adjudication, we performed corpus-wide consistency and
referential-integrity checks. The final dataset contains 98 paper
records, 284 unique atomic attack IDs, 452 unique safeguard IDs, and
660 unique attack--safeguard pairs. Every atomic attack has exactly one
primary C1--C8 class, and the class counts sum to 284. Every final
attack--safeguard relationship references a valid canonical attack and
safeguard record and contains both assessment and coverage status.

The companion artifact preserves coder-specific source IDs, canonical
record mappings, structural adjudication decisions, field-level
adjudication status, and evidence needed to trace the final synthesis
back to the reviewed sources.




%% file: sections/04-taxonomy.tex
\section{RQ1: Taxonomy of Malicious Attack Vectors Targeting Agentic AI}
\label{sec:taxonomy}

Resolver adjudication yielded 284 atomic attacks across eight
causal-mechanism classes. Table~\ref{tab:rq1-summary} summarizes the distribution of 284 adjudicated atomic attacks across eight mutually exclusive primary classes.

\par
\hspace*{1em}\textbf{Attack Class \#1 - Prompt Injection and Action Hijacking.}
The instruction-abuse class is covered by 38 papers (38.8\%) and
accounts for 57 atomic attacks (20.07\%). 
Through instruction--data confusion, malicious payloads
in webpages, emails, and retrieved results can hijack
agent planning and tool use, crossing authorization
boundaries.
\par
\hspace*{1em}\textbf{Attack Class \#2 - Tool, MCP, and Action-Layer Attacks.}
This category is covered by 21 papers (21.4\%) and accounts for
70 atomic attacks (24.65\%), the largest share of atomic attacks in the
final corpus. These attacks can manipulate semantic interfaces, including
MCP metadata, function names, parameter schemas, and tool
descriptions, without requiring direct code execution.
Runtime rug-pull attacks further demonstrate that tools may
behave benignly during inspection but change after deployment.

\par
\hspace*{1em}\textbf{Attack Class \#3 - Memory and RAG Poisoning.}
This attack class is covered by 21 papers (21.4\%) and accounts for
34 atomic attacks (11.97\%). The evidence highlights a shift in the
threat model toward delayed activation: poisoned memory or RAG content
can remain dormant across sessions and alter agent behavior when
retrieved during a later task. This persistence limits single-turn
evaluation, requiring agent safety
assessment across state transitions,
retrieval events, and task contexts.

\par
\hspace*{1em}\textbf{Attack Class \#4 - Multi-Agent Attacks.}
This attack class is covered by 12 papers (12.2\%) and accounts for
18 atomic attacks (6.34\%). Multi-agent systems introduce topology-dependent risks:
compromised outputs can become authoritative inputs to
other agents and propagate across collaborative workflows.
Even individually benign-looking messages may combine
into malicious behavior, requiring defenses to monitor
collaborative trajectories rather than isolated agents.

\par
\hspace*{1em}\textbf{Attack Class \#5 - Backdoors and Trigger-Conditioned Compromise.}
This attack class is covered by 14 papers (14.3\%) and accounts for
19 atomic attacks (6.69\%). Unlike model-weight backdoors, latent agentic backdoors
can also be implanted in skills, workflows, policies,
tools, or chat templates.
These payloads may evade static inspection and
activate only under specific environmental conditions, trigger
patterns, or stages of tool use~\cite{wang2024badagent,yang2024watch}.

\par
\hspace*{1em}\textbf{Attack Class \#6 - Computer-Use and Coding-Agent Attacks.}
This attack class is covered by 10 papers (10.2\%) and accounts for
24 atomic attacks (8.45\%). It poses direct execution risks because
model outputs can translate into browser actions, database queries,
file modifications, package installations, or shell commands. 
The impact of command injection, unsafe
execution, and data exfiltration depends
on agent privileges and execution
boundaries.

\par
\hspace*{1em}\textbf{Attack Class \#7 - Agent Skills and Supply-Chain Attacks.}
Attacks in this category are covered by 17 papers (17.3\%) and account
for 51 atomic attacks (17.96\%). Reused third-party skills, repositories, dependencies,
and remote services cross trust boundaries before agents
execute them or follow their instructions.
Provenance alone is insufficient: dynamic
skill mutation and malicious behavior
distributed across auxiliary files,
dependencies, or remote services can
evade scanning~\cite{ji2026cloak,liu2026maliciousskills,schmotz2026skillinject}.

\par
\hspace*{1em}\textbf{Attack Class \#8 - Goal, Task, and Planning Manipulation.}
This attack class is covered by four papers (4.1\%) and accounts for
11 atomic attacks (3.87\%). These attacks target autonomous goal
management and task decomposition rather than a single tool, memory
store, or execution interface. Attackers can redirect a legitimate task
through seemingly valid substeps that
collectively produce a malicious outcome,
exposing gaps in step-level safety checks.

\begin{tcolorbox}[title=\textbf{Key Findings for RQ1}]
We identified 284 atomic attacks across 98 sources and organized them into eight causal-mechanism classes. Tool, MCP, and action-layer attacks account for the largest share of atomic attacks (70, 24.65\%), while the prompt-injection and action-hijacking class
has the broadest paper coverage (38 papers, 38.8\%). Agent skills and supply-chain attacks are the third-largest class (51, 17.96\%). These frequencies reflect representation in the reviewed literature, not real-world prevalence or severity, and highlight the system-level nature of threats spanning instructions, memory, retrieval, tools, code execution, external services, and multi-agent communication.

\end{tcolorbox}

\begin{table*}[!t]
\centering
\caption{\textsc{Corpus and coding summary by final attack class.}}
\label{tab:rq1-summary}

\scriptsize
\setlength{\tabcolsep}{3.5pt}
\renewcommand{\arraystretch}{1.08}

\begin{tabularx}{\textwidth}{
    c
    >{\raggedright\arraybackslash}p{0.27\textwidth}
    >{\centering\arraybackslash}p{0.10\textwidth}
    >{\centering\arraybackslash}p{0.085\textwidth}
    >{\centering\arraybackslash}p{0.085\textwidth}
    >{\raggedright\arraybackslash}X
}
\toprule
\rowcolor{white}
\textbf{ID} &
\textbf{Attack class} &
\textbf{Papers(\%)} &
\textbf{Atomic attacks} &
\textbf{\% of attacks} &
\textbf{Representative mechanisms} \\
\midrule

\rowcolor{gray!10}
C1 &
Prompt Injection and Action Hijacking &
38 (38.8\%) &
57 &
20.07\% &
Direct/indirect prompt injection; action hijacking \\

\rowcolor{white}
C2 &
Tool, MCP, and Action-Layer Attacks &
21 (21.4\%) &
70 &
24.65\% &
Tool/MCP poisoning; metadata and invocation manipulation \\

\rowcolor{gray!10}
C3 &
Memory and RAG Poisoning &
21 (21.4\%) &
34 &
11.97\% &
Memory poisoning; RAG/retrieval poisoning \\

\rowcolor{white}
C4 &
Multi-Agent Attacks &
12 (12.2\%) &
18 &
6.34\% &
Malicious inter-agent communication; propagation \\

\rowcolor{gray!10}
C5 &
Backdoors and Trigger-Conditioned Compromise &
14 (14.3\%) &
19 &
6.69\% &
Backdoor triggers; latent conditional compromise \\

\rowcolor{white}
C6 &
Computer-Use and Coding-Agent Attacks &
10 (10.2\%) &
24 &
8.45\% &
Code/command injection; browser and shell abuse \\

\rowcolor{gray!10}
C7 &
Agent Skills and Supply-Chain Attacks &
17 (17.3\%) &
51 &
17.96\% &
Malicious skills; dependency and package compromise \\

\rowcolor{white}
C8 &
Goal, Task, and Planning Manipulation &
4 (4.1\%) &
11 &
3.87\% &
Task-state manipulation; planning and goal hijacking \\

\addlinespace[2pt]
\textbf{Total} &
&
\textbf{98} &
\textbf{284} &
\textbf{100.00\%} &
\\
\bottomrule
\end{tabularx}

\vspace{1mm}
\begin{minipage}{\textwidth}
\scriptsize
\textit{Note.}
Paper counts are non-exclusive because one paper may contain attacks
from multiple primary classes. Each atomic attack is assigned exactly
one primary class according to its causal mechanism; therefore, the
atomic attack counts sum to 284.
\end{minipage}
\end{table*}

%% file: sections/05-incidents-observations.tex
\section{RQ2: Patterns Across Attack Vectors}
\label{sec:evidence_analysis}
We identify seven non-exclusive system-level patterns
across the 284 atomic attacks; individual attacks may
exhibit multiple patterns~\cite{greshake2023notwhat,liu2023promptinjection,
schmotz2026skillinject}.

\par
\hspace*{1em}\textbf{Pattern \#1 - Instruction--Data Confusion.}
Within C1, 56 of 57 attacks (98.2\%) involve context
interpretation or instruction following, while 52 of 57
(91.2\%) activate when attacker-controlled content is
processed. When agents combine trusted instructions with
untrusted webpages, emails, documents, retrieved content,
memory entries, tool outputs, or inter-agent messages,
attackers can exploit instruction--data confusion to turn
untrusted data into operational instructions, evidence,
authority, or permission
~\cite{greshake2023notwhat,liu2023promptinjection,
schmotz2026skillinject}.

\par
\hspace*{1em}\textbf{Pattern \#2 - Delayed Activation Across Components.}
Insertion may precede activation: 31 of 34 C3
attacks (91.2\%) activate upon retrieval or recall,
while all 19 C5 attacks activate upon a trigger match.
Payloads planted in memory, retrieval corpora,
tool descriptions, dependencies, skills, MCP
servers, or webpages can activate during later
recall, tool selection, execution, installation,
or trigger events~\cite{
chen2024agentpoison,zou2025poisonedrag,wang2026mcptox,wang2024badagent,
yang2024watch}. 
Consequently, single-turn scans can miss attacks
that activate under subsequent contexts, permission settings, tool configurations, or changes in remote state~\cite{ji2026cloak}.

Table~\ref{tab:insert_activation} summarizes selected
insertion--activation pairs; the complete co-occurrence
matrix is available in the companion artifact.


\par
\hspace*{1em}\textbf{Pattern \#3 - Metadata as an Attack Surface.}
Metadata can steer agent decisions during tool, skill,
server, and package discovery or selection. In C2, 68 of
70 attacks (97.1\%) involve tool discovery, selection,
or invocation. Tool descriptions, MCP metadata, function
schemas, registry entries, skill descriptions, and model
or dataset cards can influence agent behavior without
containing executable code~\cite{ladisa2023sok,
liu2026maliciousskills,schmotz2026skillinject,wang2026mcptox}.
Accordingly, security reviews must assess natural-language descriptions
and metadata claims, not only code and binaries.

\par
\hspace*{1em}\textbf{Pattern \#4 - Capability Abuse and Side-Effect Execution.}
High-impact attacks exploit an agent's ability to act
beyond text generation. In C2, 52 of 70 attacks (74.3\%)
seek unauthorized action, while 14 of 24 C6 attacks
(58.3\%) involve code generation or execution. When
model outputs become tool calls, shell commands, browser
actions, code edits, or API requests, attacks can affect
real assets through secret exfiltration, unauthorized
operations, file modifications, or dependency
installation~\cite{greshake2023notwhat,isbarov2026agentproxy,
ji2026cloak,liu2026maliciousskills,liu2023promptinjection}.
This pattern demonstrates that tool permissions, runtime isolation, and
execution context are integral to the security model.
\par
\hspace*{1em}\textbf{Pattern \#5 - Persistence and Cross-Session Risk.}
Agentic systems maintain long-lived state through memory,
retrieval corpora, installed skills, dependencies, MCP servers,
browser sessions, and backdoors. In C3, persistent control
is an objective in 15 of 34 attacks (44.1\%). Such state
can sustain malicious influence across sessions or enable
activation during later recall, retrieval, installation,
updates, or tool use~\cite{
chen2024agentpoison,liu2026maliciousskills,schmotz2026skillinject,
wang2024badagent,yang2024watch,zou2025poisonedrag}. Safety assessment must
therefore cover state stored, reused, updated, or inherited
across tasks, not merely initial insertion.

\par
\hspace*{1em}\textbf{Pattern \#6 - Composition and Propagation Risk.}
Attacks may emerge from interactions among seemingly benign
components: tool outputs can influence subsequent calls,
poisoned retrieval can alter plans, and skill dependencies
can execute undisclosed behavior~\cite{ji2026cloak,liu2026maliciousskills,
schmotz2026skillinject,wang2026mcptox}. In multi-agent settings, one
agent's compromised output can become another agent's instruction or
evidence, allowing malicious behavior to propagate through delegation
and communication channels. 
This pattern is especially pronounced in
C4, where all 18 attacks (100\%) involve agent-to-agent propagation
~\cite{arif2026conjunctive,lee2024promptinfection}. Thus, security evaluation must trace
the complete path from attacker-controlled content to
privileged action, rather than assess components in isolation.
\par
\hspace*{1em}\textbf{Pattern \#7 - Defense Brittleness.}
Single-layer defenses remain brittle.
Prompt guards may detect obvious malicious instructions but miss
indirect, paraphrased, multi-turn, or context-dependent attacks
~\cite{greshake2023notwhat,liu2023promptinjection}. Static scanners may overlook malicious
semantic instructions, deceptive metadata, auxiliary files,
or runtime-only behavior
~\cite{ji2026cloak,liu2026maliciousskills,schmotz2026skillinject}.
Human- or LLM-based approval mechanisms can also be undermined when
agents produce misleading rationales or distribute dangerous behavior
across several benign-looking steps~\cite{isbarov2026agentproxy}.
Together, these limitations suggest that defenses must operate across
multiple stages of the agent lifecycle rather than rely on a single
inspection point.

\begin{table}[!t]
\centering
\caption{Selected insertion--activation code pairs
observed in the reviewed corpus. Counts indicate
atomic attacks exhibiting each coded pair.}
\label{tab:insert_activation}

\small
\setlength{\tabcolsep}{4pt}
\renewcommand{\arraystretch}{1.05}

\begin{tabularx}{\columnwidth}{@{}
    >{\raggedright\arraybackslash}X
    >{\raggedright\arraybackslash}X
    c
@{}}
\toprule
\textbf{Insertion point} &
\textbf{Activation condition} &
\textbf{Count} \\
\midrule

\rowcolor{gray!10}
Tool description / schema &
Tool selection or invocation &
33 \\

\rowcolor{white}
External content &
Immediate processing &
28 \\

\rowcolor{gray!10}
Skill package / file &
Code execution or build &
22 \\

\rowcolor{white}
Tool output / resource &
Immediate processing &
18 \\

\rowcolor{gray!10}
Inter-agent message &
Agent-to-agent propagation &
16 \\

\rowcolor{white}
Input prompt / context &
Immediate processing &
14 \\

\rowcolor{gray!10}
Model weights / adapter / template &
Trigger match &
13 \\

\rowcolor{white}
Skill package / file &
Installation, loading, or update &
11 \\

\rowcolor{gray!10}
Memory-write path &
Retrieval or memory recall &
10 \\

\bottomrule
\end{tabularx}
\end{table}














\begin{keyfindingbox}{Key Findings for RQ2}

We identified seven recurring patterns across the
284 atomic attacks in the reviewed corpus.
Within this corpus, activation patterns are
strongly associated with the assigned attack
classes: all 19 backdoor attacks activate on
trigger match, all 18 multi-agent attacks involve
agent-to-agent propagation, 53 of 70 tool/MCP
attacks (75.7\%) activate during tool selection
or invocation, and 31 of 34 memory/RAG attacks
(91.2\%) activate during retrieval or recall.
Some of these class-conditional patterns,
particularly trigger matching in C5 and
agent-to-agent propagation in C4, partly
reflect the operational definitions of the
corresponding classes. These findings
characterize the reviewed evidence rather
than the prevalence of these behaviors
in deployed agentic systems.
\end{keyfindingbox}


%% file: sections/06-safeguards.tex
\section{RQ3: Safeguards and Coverage Gaps}
\label{sec:safeguards}

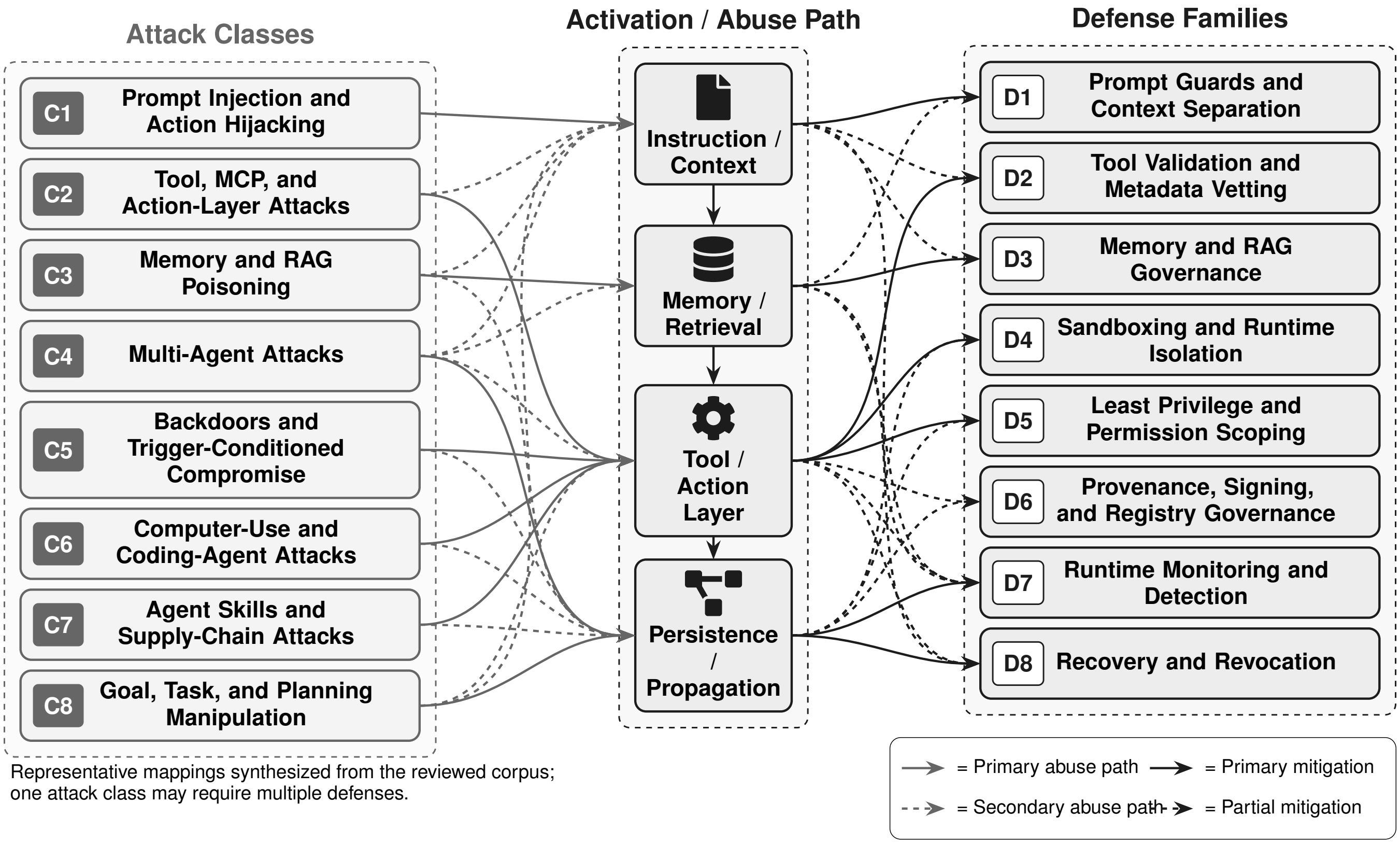
\begin{figure*}[t]
    \centering
    \begingroup
    \input{figures/attack_defense_mapping_tikz}
    \endgroup
    \caption{Attack and defense mapping for agentic AI systems.}
    \label{fig:attack_defense_mapping}
\end{figure*}

Figure~\ref{fig:attack_defense_mapping} maps eight attack
classes to their abuse paths and representative defense
families (D1--D8) distinct from the five control types in Section~\ref{sec:safeguard_coding}.
Its links indicate qualitative mitigation
relevance, not measured effectiveness or relationship-level
coverage.
Our coding identified 452 atomic safeguards and
660 explicit source-supported attack--safeguard relationships. Of these,
278 (42.1\%) were directly assessed and 382 (57.9\%) were not evaluated.
Among the assessed relationships, coverage was strong for 32 (11.5\%),
partial for 107 (38.5\%), and limited for 139 (50.0\%).

\par
\hspace*{1em}\textbf{MCP Tool Description
Poisoning (F023-A01).}
In the attack recorded as F023-A01~\cite{liu2026whenmanuallies}, an attacker embeds malicious instructions
in an MCP tool description. During tool selection,
the agent may interpret these instructions as
operational requirements and perform unauthorized
actions. The decisive mechanism is manipulation
of the tool interface (C2), rather than generic
instruction hijacking (C1).

In Fig.~\ref{fig:attack_defense_mapping},
this attack follows the tool/action abuse path.
Metadata validation (D2) and runtime monitoring
(D7) are relevant defense families, but their
relevance does not establish effectiveness.
The source evaluates LlamaGuard-based filtering
(F023-S01, limited coverage) and proposes
reactive self-correction (F023-S02, partial
coverage in our coding). The latter may
mitigate consequences after execution rather
than prevent the initial compromise.

\par
\hspace*{1em}\textbf{Memory/RAG
Poisoning (F031-A01).}
AgentPoison~\cite{chen2024agentpoison} inserts
attacker-controlled content into an agent's
retrieval corpus. When a matching query
retrieves the poisoned content, the agent
may deviate from its intended task.
The decisive mechanism is compromised
retrieval state (C3), rather than ordinary
prompt injection (C1). In
Fig.~\ref{fig:attack_defense_mapping},
the attack follows the memory/retrieval path,
where memory governance (D3) and detection
(D7) are relevant. However, the evaluated
Perplexity Filter (F031-S01) and Query
Rephrasing (F031-S02) both showed limited
coverage under the reported conditions.

\par
\hspace*{1em}\textbf{Trigger-Conditioned
Backdoor (F001-A01).}
BadAgent~\cite{wang2024badagent} implants
a backdoor through model fine-tuning.
When the trigger appears, the compromised
agent can perform attacker-desired actions,
including unsafe tool use. The primary
mechanism is trigger-conditioned compromise
(C5), not manipulation of the tool interface
(C2). In Fig.~\ref{fig:attack_defense_mapping},
the attack spans trigger activation and
the tool/action path. Detection (D7) and
recovery (D8) are relevant defense families,
but their effectiveness is not established
by the mapping. The source evaluates
clean-data fine-tuning (F001-S01), which
provides limited coverage in our coding.

The mapping is many-to-many: attack classes may involve multiple abuse
paths and require complementary defenses. Prompt-level defenses address
some forms of prompt injection and action hijacking but are less effective
against indirect, multi-turn, tool-mediated, or persistent attacks
~\cite{greshake2023notwhat,isbarov2026agentproxy,
liu2023promptinjection}. Tool validation and runtime permissions are more
relevant to tool, MCP, computer-use, and coding-agent attacks, whereas
provenance and registry governance can mitigate supply-chain risks without
establishing that a signed artifact, model, skill, or dependency is
behaviorally benign~\cite{bennet2024aibom,gebru2021datasheets,
ladisa2023sok,mitchell2019modelcards,zhao2024models}.
\par
\hspace*{1em}\textbf{Safeguard \#1 - Prevention and Detection.} 
Preventive safeguards aim to block malicious content, tools, or artifacts
from entering the workflow or being selected for use. These measures
include context separation, instruction hierarchies, tool allowlists,
metadata validation, provenance checks, dependency pinning, and registry
governance~\cite{greshake2023notwhat,liu2023promptinjection,
wang2026mcptox,ladisa2023sok,zhao2024models}. Preventive controls are
the most common safeguard type in our coding, accounting for 247 of 452
atomic safeguards (54.6\%). Detective safeguards identify suspicious
content or behavior through prompt-injection detectors, static and
semantic scanners, tool-call inspection, runtime tracing, and audit
logging~\cite{ji2026cloak,liu2026maliciousskills,
schmotz2026skillinject}, and account for 141 of 452 atomic safeguards
(31.2\%). Although useful, these controls remain incomplete: prompt
guards struggle with indirect, paraphrased, multi-turn, and
context-dependent attacks, while static scanners may miss semantically
malicious instructions, deceptive metadata, auxiliary files, or
runtime-only behavior.
\par
\hspace*{1em}\textbf{Safeguard \#2 - Containment and Least Privilege.} 
Containment safeguards address cases in which preventive and detective
controls fail. They limit impact through sandboxing, least privilege,
scoped credentials, network and filesystem restrictions, and human
approval for sensitive actions~\cite{chhabra2026agenticsecurity,
dehghantanha2026attacksurface,kim2026attackdefenselandscape}.
Containment controls account for 48 of 452 atomic safeguards (10.6\%).
These controls are especially important for tool, MCP, computer-use,
coding-agent, and supply-chain attacks because the potential harm depends
on the files, credentials, APIs, commands, and network resources available
to the agent~\cite{isbarov2026agentproxy,liu2026maliciousskills,
wang2026mcptox}. However, sandboxing and approval gates do not prevent
malicious content from entering the agent's context; they only limit the
damage that can result from its interpretation or execution.
\par
\hspace*{1em}\textbf{Safeguard \#3: Recovery and Remaining Gaps.} 
Corrective and recovery safeguards address compromised state after
activation. They include memory cleanup, poisoned-corpus removal,
dependency rollback, artifact revocation, credential rotation, downstream
notification, and incident response~\cite{bennet2024aibom,
chen2024agentpoison,ladisa2023sok,zou2025poisonedrag}. These controls
are comparatively rare in the reviewed literature: corrective safeguards
account for 14 of 452 atomic safeguards (3.1\%), while recovery
safeguards account for only 2 of 452 (0.4\%).
Recovery is especially relevant to memory and RAG attacks, malicious
skills, supply-chain attacks, and backdoors because compromised state may
persist beyond the initial insertion and spread when an artifact is
copied, installed, or embedded in another workflow~\cite{ji2026cloak,
liu2026maliciousskills,schmotz2026skillinject,wang2024badagent,
yang2024watch}. More broadly, substantial evaluation gaps remain:
382 of 660 explicit attack--safeguard relationships (57.9\%) were not
evaluated. Among the 278 directly assessed relationships, only
32 (11.5\%) were classified as providing strong coverage, compared with 107 (38.5\%)
partial and 139 (50.0\%) limited coverage. These results indicate that
many proposed defenses remain either untested or only partially effective
against the attacks they are intended to address.
\begin{keyfindingbox}{Key Findings for RQ3}
Within the reviewed corpus, no single safeguard
was shown to provide comprehensive protection
across all attack classes. Effective defense
therefore requires complementary preventive,
detective, containment, corrective, and recovery
controls. However, 382 of 660 explicit
attack--safeguard relationships (57.9\%) were
not evaluated, and only 32 of the 278 directly
assessed relationships (11.5\%) showed strong
coverage. These findings reflect the available
evaluation evidence rather than universal
safeguard effectiveness. Important evidence gaps
remain in delayed, semantic, and compositional
attacks, as well as post-compromise recovery.

\end{keyfindingbox}

%% file: figures/attack_defense_mapping_tikz.tex
\colorlet{mapattackred}{attackred!85!black}
\colorlet{mapattackredbg}{attackred!8}
\colorlet{mappathblue}{sysblue!80!black}
\colorlet{mappathbluebg}{sysblue!8}
\colorlet{mapdefensegreen}{sysblue!80!black}
\colorlet{mapdefensegreenbg}{sysblue!8}
\colorlet{mappurplearrow}{attackred!85!black}

\newcommand{\attackmapnode}[4][]{%
    \node[mapattackbox,#1] (#2) {%
        \begin{minipage}[c]{1cm}
            \centering\tikz[baseline=(badge.base)]{\node[mapclabel] (badge) {#2};}
        \end{minipage}\hspace{2mm}%
        \begin{minipage}[c]{5.2cm}
            \centering
            \textbf{#3}\par
            \if\relax\detokenize{#4}\relax
            \else
                \vspace{1mm}%
                {\raggedright\mdseries\normalsize\textcolor{black}{#4}\par}%
            \fi
        \end{minipage}%
    };%
}

\newcommand{\defensemapnode}[4][]{%
    \node[mapdefensebox,#1] (#2) {%
        \begin{minipage}[c]{1cm}
            \centering\tikz[baseline=(badge.base)]{\node[mapdlabel] (badge) {#2};}
        \end{minipage}\hspace{2mm}%
        \begin{minipage}[c]{5.2cm}
            \centering
            \textbf{#3}\par
            \if\relax\detokenize{#4}\relax
            \else
                \vspace{1mm}%
                {\raggedright\mdseries\normalsize\textcolor{black}{#4}\par}%
            \fi
        \end{minipage}%
    };%
}

\resizebox{\textwidth}{!}{%
\begin{tikzpicture}[
    >={Stealth[scale=1.2]},
    every node/.append style={font=\sffamily},
    mapattackbox/.style={rectangle,rounded corners=2mm,draw=mapattackred,
        very thick,fill=mapattackredbg,text width=7cm,align=left,inner sep=2mm,
        outer sep=0,minimum height=1.3cm,font=\sffamily\large},
    mappathbox/.style={rectangle,rounded corners=2mm,draw=mappathblue,
        very thick,fill=mappathbluebg,text width=2.5cm,align=center,inner sep=2mm,
        outer sep=0,minimum height=1.3cm,font=\sffamily\large},
    mapdefensebox/.style={rectangle,rounded corners=2mm,draw=mapdefensegreen,
        very thick,fill=mapdefensegreenbg,text width=7cm,align=left,inner sep=2mm,
        outer sep=0,minimum height=1.3cm,font=\sffamily\large},
    mapclabel/.style={rectangle,rounded corners=1mm,fill=mapattackred,
        text=white,font=\sffamily\large\bfseries,minimum width=0.8cm,minimum height=0.8cm,
        align=center},
    mapdlabel/.style={rectangle,rounded corners=1mm,fill=white,
        draw=mapdefensegreen,thick,text=mapdefensegreen,font=\sffamily\large\bfseries,
        minimum width=0.8cm,minimum height=0.8cm,align=center},
    mapattackcontainer/.style={rectangle,draw=mapattackred,dashed,
        rounded corners=2mm,inner sep=3mm,fill=attackred!3,thick},
    mappathcontainer/.style={rectangle,draw=mappathblue,dashed,
        rounded corners=2mm,inner sep=3mm,fill=sysblue!3,thick},
    mapdefensecontainer/.style={rectangle,draw=mapdefensegreen,dashed,
        rounded corners=2mm,inner sep=3mm,fill=sysblue!3,thick},
    mapprimaryattack/.style={-{Stealth[scale=1.2]},draw=mapattackred,very thick},
    mapsecondaryattack/.style={-{Stealth[scale=1.2]},draw=mappurplearrow,very thick,dashed},
    mapprimarydefense/.style={-{Stealth[scale=1.2]},draw=mapdefensegreen,very thick},
    mappartialdefense/.style={-{Stealth[scale=1.2]},draw=mapdefensegreen,very thick,dashed},
    mappathflow/.style={-{Stealth[scale=1.2]},draw=mappathblue,very thick}
]

\attackmapnode{C1}{Prompt Injection and Action Hijacking}
    {}
\attackmapnode[below=0.2cm of C1]{C2}{Tool, MCP, and Action-Layer Attacks}
    {}
\attackmapnode[below=0.2cm of C2]{C3}{Memory and RAG Poisoning}
    {}
\attackmapnode[below=0.2cm of C3]{C4}{Multi-Agent Attacks}
    {}
\attackmapnode[below=0.2cm of C4]{C5}{Backdoors and Trigger-Conditioned Compromise}
    {}

\attackmapnode[below=0.2cm of C5]{C6}{Computer-Use and Coding-Agent Attacks}
    {}
\attackmapnode[below=0.2cm of C6]{C7}{Agent Skills and Supply-Chain Attacks}
    {}
\attackmapnode[below=0.2cm of C7]{C8}{Goal, Task, and Planning Manipulation}
    {}

\begin{pgfonlayer}{background}
    \node[mapattackcontainer,fit=(C1) (C8)] (AttackGroup) {};
\end{pgfonlayer}
\node[above=0.2cm of AttackGroup,font=\sffamily\Large\bfseries\color{mapattackred}]
    {Attack Classes};

\node[mappathbox,right=4cm of C1,yshift=-0.2cm] (P1)
    {\textcolor{mappathblue}{\Huge\faFile}\\[1mm]\textbf{Instruction /\\ Context}};
\node[mappathbox,right=4cm of C3,yshift=-0.2cm] (P2)
    {\textcolor{mappathblue}{\Huge\faDatabase}\\[1mm]\textbf{Memory /\\ Retrieval}};
\node[mappathbox,right=4cm of C5,yshift=-0.2cm] (P3)
    {\textcolor{mappathblue}{\Huge\faCog}\\[1mm]\textbf{Tool /\\ Action Layer}};
\node[mappathbox,right=4cm of C7,yshift=-0.2cm] (P4)
    {\textcolor{mappathblue}{\Huge\faProjectDiagram}\\[1mm]\textbf{Persistence /\\ Propagation}};

\begin{pgfonlayer}{background}
    \node[mappathcontainer,fit=(P1) (P4)] (PathGroup) {};
\end{pgfonlayer}
\node[above=0.2cm of PathGroup,font=\sffamily\Large\bfseries\color{mappathblue}]
    {Activation / Abuse Path};

\draw[mappathflow] (P1) -- (P2);
\draw[mappathflow] (P2) -- (P3);
\draw[mappathflow] (P3) -- (P4);

\defensemapnode[right=3.5cm of P1,yshift=0.5cm]{D1}
    {Prompt Guards and Context Separation}
    {}
\defensemapnode[below=0.2cm of D1]{D2}{Tool Validation and Metadata Vetting}
    {}
\defensemapnode[below=0.2cm of D2]{D3}{Memory and RAG Governance}
    {}
\defensemapnode[below=0.2cm of D3]{D4}{Sandboxing and Runtime Isolation}
    {}
\defensemapnode[below=0.2cm of D4]{D5}{Least Privilege and Permission Scoping}
    {}
\defensemapnode[below=0.2cm of D5]{D6}{Provenance, Signing, and Registry Governance}
    {}
\defensemapnode[below=0.2cm of D6]{D7}{Runtime Monitoring and Detection}
    {}
\defensemapnode[below=0.2cm of D7]{D8}{Recovery and Revocation}
    {}

\begin{pgfonlayer}{background}
    \node[mapdefensecontainer,fit=(D1) (D8)] (DefenseGroup) {};
\end{pgfonlayer}
\node[above=0.2cm of DefenseGroup,font=\sffamily\Large\bfseries\color{mapdefensegreen}]
    {Defense Families};

\draw[mapprimaryattack] (C1.east) -- (P1.west);
\draw[mapsecondaryattack] (C2.east) to[out=0,in=180] (P1.west);
\draw[mapprimaryattack] (C2.east) to[out=0,in=180] (P3.west);
\draw[mapsecondaryattack] (C3.east) to[out=0,in=180] (P1.west);
\draw[mapprimaryattack] (C3.east) -- (P2.west);
\draw[mapsecondaryattack] (C3.east) to[out=0,in=180] (P4.west);
\draw[mapsecondaryattack] (C4.east) to[out=0,in=180] (P1.west);
\draw[mapsecondaryattack] (C4.east) to[out=0,in=180] (P2.west);
\draw[mapsecondaryattack] (C4.east) to[out=0,in=180] (P3.west);
\draw[mapprimaryattack] (C4.east) to[out=0,in=180] (P4.west);
\draw[mapprimaryattack] (C5.east) to[out=0,in=180] (P3.west);
\draw[mapsecondaryattack] (C5.east) to[out=0,in=180] (P4.west);
\draw[mapprimaryattack] (C6.east) to[out=0,in=180] (P3.west);
\draw[mapsecondaryattack] (C6.east) to[out=0,in=180] (P4.west);
\draw[mapprimaryattack] (C7.east) to[out=0,in=180] (P3.west);
\draw[mapsecondaryattack] (C7.east) to[out=0,in=180] (P4.west);
\draw[mapsecondaryattack] (C8.east) to[out=0,in=180] (P1.west);
\draw[mapsecondaryattack] (C8.east) to[out=0,in=180] (P3.west);
\draw[mapprimaryattack] (C8.east) to[out=0,in=180] (P4.west);

\draw[mapprimarydefense] (P1.east) to[out=0,in=180] (D1.west);
\draw[mappartialdefense] (P1.east) to[out=0,in=180] (D2.west);
\draw[mappartialdefense] (P1.east) to[out=0,in=180] (D3.west);
\draw[mappartialdefense] (P1.east) to[out=0,in=180] (D7.west);
\draw[mappartialdefense] (P2.east) to[out=0,in=180] (D1.west);
\draw[mapprimarydefense] (P2.east) to[out=0,in=180] (D3.west);
\draw[mappartialdefense] (P2.east) to[out=0,in=180] (D7.west);
\draw[mapprimarydefense] (P3.east) to[out=0,in=180] (D2.west);
\draw[mapprimarydefense] (P3.east) to[out=0,in=180] (D4.west);
\draw[mapprimarydefense] (P3.east) to[out=0,in=180] (D5.west);
\draw[mappartialdefense] (P3.east) to[out=0,in=180] (D6.west);
\draw[mappartialdefense] (P3.east) to[out=0,in=180] (D7.west);
\draw[mappartialdefense] (P4.east) to[out=0,in=180] (D4.west);
\draw[mappartialdefense] (P4.east) to[out=0,in=180] (D5.west);
\draw[mappartialdefense] (P4.east) to[out=0,in=180] (D6.west);
\draw[mapprimarydefense] (P4.east) to[out=0,in=180] (D7.west);
\draw[mapprimarydefense] (P4.east) to[out=0,in=180] (D8.west);
\draw[mappartialdefense] (P2.east) to[out=0,in=180] (D8.west);
\draw[mappartialdefense] (P3.east) to[out=0,in=180] (D8.west);

\node[align=left,font=\sffamily\normalsize,anchor=north west]
    at (AttackGroup.south west) {Representative mappings synthesized from the reviewed corpus;\\
    one attack class may require multiple defenses.};

\node[draw=black,rounded corners=2mm,inner sep=2mm,anchor=north east]
    at ([yshift=-4mm]DefenseGroup.south east) {%
    \begin{tikzpicture}
        \draw[mapprimaryattack] (0,1.0) -- (0.8,1.0)
            node[right,text=black,font=\sffamily\normalsize] {= Primary abuse path};
        \draw[mapsecondaryattack] (0,0.25) -- (0.8,0.25)
            node[right,text=black,font=\sffamily\normalsize] {= Secondary abuse path};
        \draw[mapprimarydefense] (4.6,1.0) -- (5.4,1.0)
            node[right,text=black,font=\sffamily\normalsize] {= Primary mitigation};
        \draw[mappartialdefense] (4.6,0.25) -- (5.4,0.25)
            node[right,text=black,font=\sffamily\normalsize] {= Partial mitigation};
    \end{tikzpicture}%
};

\end{tikzpicture}%
}

%% file: sections/07-research-gaps.tex
\section{Implications and Research Directions}
\label{sec:research_agenda}

Our taxonomy and safeguard mapping reveal open problems
spanning context, retrieval, memory, tools, execution,
external services, and multi-agent workflows. Because
attacks can cross components, mutate, and propagate,
securing agentic AI systems requires lifecycle-aware
analysis beyond isolated prompts, artifacts, or tools.

\par
\hspace*{1em}\textbf{Implication \#1 - Standardized Reporting for Agentic-AI Attack Vectors.}
The agentic-AI security literature uses overlapping terminology, including
prompt injection~\cite{liu2023promptinjection,greshake2023notwhat},
jailbreak~\cite{Zou2023UniversalAdversarial},
poisoning~\cite{Carlini2023PoisoningLLMs},
backdoor~\cite{Xu2024BackdoorLLM},
trojan~\cite{Kurita2020TrojanLLMs}, malicious
skill~\cite{schmotz2026skillinject}, tool
poisoning~\cite{Sneh2025ToolTweak}, and supply-chain
attack~\cite{OWASP2025SupplyChain}. These terms conflate carriers (prompts, tool descriptions),
causal mechanisms (poisoning, hijacking), and effects
(exfiltration, execution, persistence), hindering
cross-study comparison and defense development~\cite{Zhang2025AgentSecurityBench,
Deng2026LLMAgentSecurity}.

We propose a standardized reporting schema that explicitly separates the
primary causal mechanism from orthogonal attack dimensions, including the
carrier, delivery channel, insertion point, activation condition, abused
agentic capability~\cite{He2024SecurityAIAgents}, attacker objective,
downstream impact, affected stakeholder, evidence level, and evaluated
safeguards~\cite{OWASP2025PromptInjection}. Consistent reporting would support cross-study meta-analysis,
reduce attack-label ambiguity, and enable targeted defenses.

\par
\hspace*{1em}\textbf{Implication \#2 - Representative Benchmarks and Evaluation Corpora.}
Existing benchmarks often target individual attack families,
such as prompt injection or unsafe tool invocation~\cite{Debenedetti2024AgentDojo,
Xie2025ToolSafety,Zhan2024InjecAgent}, although ASB integrates multiple
attack and defense classes
~\cite{Zhang2025AgentSecurityBench}. Future benchmarks should cover realistic workflows involving
retrieval, memory, tools, execution, dependencies, external
services, and multi-agent interactions. Evaluation corpora
should include prompts, tool and skill metadata, memory and
retrieval state, dependencies, execution traces, and expected
outcomes. They should also vary permissions and environment
states, as attacks harmless in restricted sandboxes may
cause damage under broader access~\cite{Debenedetti2024AgentDojo,OWASP2025PromptInjection}.

\par
\hspace*{1em}\textbf{Implication \#3 - Cross-Component and Lifecycle Analysis.}
Agentic attacks frequently separate insertion from activation, with payloads introduced in one component but triggered
during later retrieval, tool use, installation, or execution
~\cite{chen2024agentpoison,Srivastava2025MemoryGraft,
krishna2025phantoms,liu2026maliciousskills,schmotz2026skillinject}.
This lifecycle structure requires cross-component analysis rather than
isolated artifact inspection. Effective detectors should correlate
evidence across context, memory, tool metadata and calls, dependency
changes, runtime events, and inter-agent communication
~\cite{Shin2025WorkflowProvenance,Ouyang2026MemLineage}. Such correlation
is necessary to reconstruct the path from attacker-controlled input to
security-relevant action~\cite{Leong2026ForensicTrajectory}.

\par
\hspace*{1em}\textbf{Implication \#4 - Semantic Instructions, Tool Metadata, and Skill Vetting.}
Natural-language artifacts pose a distinct challenge because they can
induce security-relevant behavior without containing conventional
malware~\cite{Jamshidi2025MCP-Guard}. A skill file, for example, may
resemble documentation yet be interpreted as an operational instruction.
Future work should therefore develop semantic vetting methods that
distinguish documentation from actionable instructions, identify hidden
or indirect requests for unsafe behavior, and detect permissions that
exceed an artifact's stated purpose~\cite{Xie2025ToolSafety,
Jamshidi2025MCP-Guard}. Such analysis should also be permission-aware,
because the risk of an instruction depends on the tools, data, credentials,
and external services available to the agent~\cite{OWASP2025SupplyChain}.

\par
\hspace*{1em}\textbf{Implication \#5 - Mutable Tools, Dependencies, and Time-Dependent Risk.}
Agentic workflows depend on resources whose behavior or content may
change after initial review, including tools, MCP servers, dependencies,
retrieval corpora, hosted services, and skill resources
~\cite{Rahman2025TestBeforeDeploy}. A component that appears benign
during evaluation may later change in behavior, ownership, metadata, or
dependency resolution~\cite{Latendresse2026SkillDependency}. This
time-dependent risk motivates immutable references, dependency snapshots,
continuous re-verification, semantic drift detection, update policies,
and downstream notification~\cite{OWASP2025SupplyChain}. 
These controls should extend beyond code packages to
mutable tools, services, and agent resources~\cite{wang2026mcptox}.

\par
\hspace*{1em}\textbf{Implication \#6 - Provenance Beyond Authenticity.}
Provenance mechanisms such as signatures, hashes,
and attestations can establish origin and integrity,
but these properties alone do not guarantee
behavioral safety~\cite{jiang2026soktaxonomyattackvectors,
liu2026maliciousskills}. A signed tool
may still contain a malicious description, and a verified skill may still
instruct an agent to exfiltrate data.
Future provenance systems for agentic AI should link authenticity evidence
to behavioral claims~\cite{Agentic2025DefenseSupplyChain}. Relevant
metadata may include capability manifests, permission requirements,
expected tool behavior, evaluated prompts and trigger conditions, sandbox
traces, dependency snapshots, and policy
attestations~\cite{Shin2025WorkflowProvenance}. Trust decisions should
also be revocable when new evidence indicates that an artifact, tool,
memory entry, or dependency is unsafe~\cite{Zhang2025MemorySurvey,
Ouyang2026MemLineage,Agentic2025DefenseSupplyChain}.

\par
\hspace*{1em}\textbf{Implication \#7 - Recovery, Revocation, and Post-Compromise Measurement.}
The reviewed safeguards are heavily concentrated on prevention,
detection, and containment: preventive controls account for 247 of 452
atomic safeguards (54.6\%), detective controls for 141 (31.2\%), and
containment controls for 48 (10.6\%), whereas corrective and recovery
safeguards account for only 14 (3.1\%) and 2 (0.4\%), respectively.
This imbalance highlights recovery as an underdeveloped area of
agentic-AI security research.

Agentic attacks can leave persistent downstream state in memory,
retrieval stores, dependencies, repositories, credentials, or connected
agents~\cite{Zhang2025MemorySurvey,Xiong2026CrossSessionInjection}.
Removing a malicious artifact may leave
downstream effects unresolved, making
recovery a first-class research priority.

Important directions include memory and poisoned-state cleanup,
dependency rollback, credential rotation, artifact revocation, forensic reconstruction, and downstream notification
~\cite{Zhang2025MemorySurvey,Leong2026ForensicTrajectory,
Ouyang2026MemLineage}. Ecosystem measurement should quantify
malicious-component availability,
propagation, and time to detection
and remediation~\cite{Koi2026ClawHavoc,Latendresse2026SkillDependency}.

%% file: sections/08-related-work.tex
\section{Related Work}
\label{sec:related}

\subsection{Prompt Injection and Agent Security}
\label{sec:related_prompt_injection}
Research on prompt injection and jailbreaks demonstrates how adversarial
instructions can override system policies, redirect model behavior, or
produce unsafe responses. LLM-integrated applications often blur the
boundary between data and instructions, allowing malicious prompts
embedded in external content to influence application
behavior~\cite{greshake2023notwhat}. Studies of deployed applications have
also demonstrated practical prompt-injection attacks across real
systems~\cite{liu2023promptinjection}. This work is foundational to agentic AI security because agents often
process external content, retrieved passages, tool outputs, memory entries,
and user instructions within a shared context. However, prompt-injection
research often emphasizes model responses or policy compliance. In agentic
systems, the potential impact is greater because injected instructions can
trigger tool calls, memory writes, file access, code execution, API
requests, or inter-agent communication. We therefore group prompt injection and
action hijacking by their shared mechanism
and assess their downstream agentic effects.

\subsection{Tool, Plugin, and MCP Security}
\label{sec:related_tool_mcp}
Research on tool, plugin, function-calling, and MCP security examines how
agents discover tools, interpret descriptions, generate parameters, call
APIs, and consume tool outputs. MCPTox shows that malicious instructions
embedded in MCP metadata can systematically compromise agent
behavior~\cite{wang2026mcptox}. Prior studies often treat tool misuse or
MCP poisoning as standalone problems. Our paper instead places tool and
MCP attacks within a broader taxonomy that compares them with memory
poisoning, malicious skills, browser and coding-agent attacks, multi-agent
propagation, and persistent compromise. Memory, retrieval-augmented generation, and multi-agent systems create
persistent and compositional attack surfaces. Zou et
al.~\cite{zou2025poisonedrag} show that malicious content injected into a
knowledge database can steer subsequent retrieval-augmented outputs.
Prompt infection and conjunctive prompt attacks further show that
malicious behavior can propagate or activate only when messages,
templates, or routes are combined across agents~\cite{arif2026conjunctive,
lee2024promptinfection}. 
Our lifecycle view distinguishes insertion,
selection, interpretation, activation,
persistence, and propagation across
these mechanisms.

\begin{table*}[!t]
\centering
\caption{\textsc{Comparison with the closest recent systematizations
of agentic-AI security.}}
\label{tab:closest_soks}
\small
\setlength{\tabcolsep}{3pt}
\renewcommand{\arraystretch}{1.02}

\begin{tabularx}{\textwidth}{@{}
    >{\raggedright\arraybackslash}p{2.0cm}
    >{\raggedright\arraybackslash}p{2.1cm}
    >{\raggedright\arraybackslash}X
    >{\raggedright\arraybackslash}X
    >{\raggedright\arraybackslash}X
@{}}
\toprule
\rowcolor{white}
\textbf{Study} &
\textbf{Primary scope} &
\textbf{Organizing lens} &
\textbf{Evidence representation} &
\textbf{Defense analysis} \\
\midrule

\rowcolor{gray!10}
Kim et al.~\cite{kim2026attackdefenselandscape}
&
General agentic-AI security
&
Agent design, threat models, and security-risk categories
&
Broad review of attack and defense methods
&
Runtime protection, secure-by-design, and component hardening
\\

\rowcolor{white}
Jiang et al.~\cite{jiang2026soktaxonomyattackvectors}
&
Agentic supply-chain runtime
&
Data/tool supply-chain attacks across runtime stages
&
Inference-time dependencies and attack-path synthesis
&
Zero-trust runtime and supply-chain defense strategies
\\

\rowcolor{gray!10}
\textbf{This work}
&
\textbf{Cross-mechanism agentic-AI attacks}
&
\textbf{284 atomic attacks; primary causal mechanisms and
orthogonal lifecycle dimensions}
&
\textbf{Source-supported atomic records: insertion,
activation, capability, objective, persistence, and evidence}
&
\textbf{452 safeguards; 660 explicit attack--safeguard
relationships with attack-specific coverage}
\\

\bottomrule
\end{tabularx}
\end{table*}

\subsection{Backdoors, Data Poisoning, and Persistent Compromise}
\label{sec:related_backdoors_poisoning}
Research on backdoors and data poisoning shows how malicious behavior can
be introduced through training, fine-tuning, adaptation, or data
collection. Surveys of training-data poisoning provide a broad view of
poisoning mechanisms and defenses in machine learning
systems~\cite{cina2022wild}. Studies of model supply-chain poisoning show
that backdoored pretrained models can persist through downstream
adaptation and transfer across tasks~\cite{wang2025model}. These attacks
are closely related to agentic AI because a backdoored model, adapter,
skill, memory entry, or workflow can remain dormant until a specific
prompt, context, or tool-use stage is encountered. In agentic systems, a
backdoor may affect not only classification or text generation but also
planning, tool invocation, command execution, memory modification, or data
exfiltration~\cite{wang2024badagent,yang2024watch}. We therefore emphasize agentic activation paths
rather than model-training mechanisms alone.
\subsection{AI Supply Chains, Malicious Skills, and Provenance}
\label{sec:related_supply_chain}
Software and AI supply-chain research examines how compromised
components cross trust boundaries through repositories, package
managers, model hubs, dataset hubs, registries, and update mechanisms
~\cite{ladisa2023sok,zhao2024models}. Agent skills extend these risks
to natural-language and tool-mediated artifacts. Recent research on
malicious agent skills shows that attacks can be concealed in skill
instructions, companion scripts, or context-dependent payloads
~\cite{liu2026maliciousskills,schmotz2026skillinject}. Provenance
mechanisms, including model cards, dataset datasheets, AI Bills of
Materials, signatures, hashes, lineage records, and attestations,
improve transparency but do not independently establish behavioral
safety~\cite{bennet2024aibom,gebru2021datasheets,
mitchell2019modelcards}. 
We treat supply chains as delivery channels
and trust boundaries, while classifying
attacks by causal mechanism and retaining
delivery, insertion, activation, capability,
objective, and impact as orthogonal dimensions.

\subsection{Comparison with Closest Surveys and SoKs}
The distinction is primarily analytical rather than topical.
Kim et al.~\cite{kim2026attackdefenselandscape} provide a broad
systems-level view of agent design, risks, and defenses, while
Jiang et al.~\cite{jiang2026soktaxonomyattackvectors} focus on
inference-time agentic supply chains. Our analysis instead uses the
atomic attack as the primary unit, separates causal class from lifecycle
and behavioral dimensions, and evaluates defenses through explicit
attack--safeguard relationships rather than only through defense-family
membership.
 
Table~\ref{tab:closest_soks} summarizes
these distinctions.

\begin{keyfindingbox}{Summary}
Unlike broad surveys or single-mechanism studies, our SoK
systematizes atomic attacks by primary causal mechanism
while retaining orthogonal lifecycle and behavioral
dimensions. It further maps explicit, source-supported
attack--safeguard relationships and attack-specific
coverage evidence.
\end{keyfindingbox}

%% file: sections/09-threats-ethics.tex
\section{Threats to Validity and Ethics}
\label{sec:threats}
Our study is subject to limitations in search coverage, publication bias, and inconsistent terminology. Relevant attacks may appear under labels such as prompt injection, tool or MCP poisoning, memory poisoning, malicious skills, backdoors, jailbreaks, or supply-chain compromise. We mitigated these limitations through multiple query families, snowballing, and grey-literature searches. However, because search results and grey-literature availability change over time, the corpus should not be interpreted as exhaustive or exactly reproducible.

The corpus combines peer-reviewed publications, preprints, and technically attributable grey literature with differing review standards and evidence availability. It also includes attacks observed in the wild, demonstrated in sandbox settings, or described theoretically; we preserve these evidence distinctions rather than treating all attacks as equivalent. Coding requires judgment because attacks may span multiple components and effects. We therefore assign each atomic attack to the class that best captures its primary causal mechanism while retaining delivery, insertion, activation, capability, objective, impact, and related properties as orthogonal dimensions.

Inter-coder agreement varied across coding dimensions, particularly for some evidence and safeguard assignments. Adjudication established a consistent final dataset but does not eliminate uncertainty in independent coding, and agreement on aligned records does not capture disagreements in record discovery or atomic split/merge decisions.

Reported frequencies reflect representation in the reviewed literature, not real-world prevalence, incidence, or risk. Multiple atomic attacks may also originate from the same source and should not be treated as independent observations.

Finally, the study reflects the ecosystem as of July 2026, and subsequent developments may affect its conclusions. Given the dual-use nature of the topic, we focus on taxonomy, lifecycle patterns, and safeguards rather than operational payloads, active endpoints, credentials, or step-by-step exploitation instructions.
%

%% file: sections/10-conclusion.tex
\section{Conclusion}
\label{sec:conclusion}
This paper systematizes malicious attacks on agentic AI systems across 98 sources, identifying 284 atomic attacks, 452 atomic safeguards, and 660 explicit source-supported attack--safeguard relationships. We organize attacks into eight causal-mechanism classes and show that agentic-AI threats are system-level, spanning prompts, memory and retrieval, tools and MCP interfaces, execution environments, dependencies, external services, and multi-agent communication.

Our analysis of seven recurring patterns also reveals uneven safeguard coverage: only 278 of 660 relationships (42.1\%) were directly assessed, and only 32 of those (11.5\%) showed strong coverage. Defensive techniques such as prompt guards, tool
validation, sandboxing, least privilege, memory controls, provenance,
and registry governance each address part of the attack surface, but
significant gaps remain in delayed, semantic, compositional, and
post-compromise settings. These findings motivate a defense-in-depth
approach to agentic AI security and provide a foundation for threat
modeling, benchmark design, tool governance, and future research.

To support transparency and reproducibility, the complete
review artifacts are available on Zenodo at
\url{https://zenodo.org/records/24009392}.

%% file: sections/appendix.tex
\section{Full Final Codebook}
\label{app:full-codebook}

Table~\ref{tab:full_final_codebook} reports the complete set of coding
dimensions used in the final analysis together with their definitions
and representative coded examples. The examples illustrate how the
coding dimensions were applied and are not verbatim quotations from the
reviewed sources.

\begin{table*}[t]
\centering
\caption{\textsc{Full final codebook used in the study. Examples illustrate
representative coded instances rather than verbatim source quotations.}}
\label{tab:full_final_codebook}
\small
\setlength{\tabcolsep}{3.5pt}
\renewcommand{\arraystretch}{1.08}

\begin{tabularx}{\textwidth}{
    L{2.8cm}
    L{1.8cm}
    >{\raggedright\arraybackslash}X
    >{\raggedright\arraybackslash}X}
\toprule
\rowcolor{white}
\textbf{Code group} &
\textbf{Dimension} &
\textbf{Description} &
\textbf{Representative example} \\
\midrule

\textbf{\textsc{Attack classification}}
& Primary class
& Assigns each atomic attack to one of eight classes according to its
primary causal mechanism.
& A malicious MCP tool description that steers an agent toward an
attacker-controlled action is coded as C2: Tool, MCP, and Action-Layer
Attacks. \\

& Causal mechanism
& Captures the causal process through which attacker-controlled content,
state, or behavior compromises the agentic workflow.
& Instructions embedded in an external document are interpreted as
operational commands: indirect prompt injection. \\

\addlinespace[4pt]

\rowcolor{gray!10}
\textbf{\textsc{Attack path}}
& Target component
& Identifies the component, state, or trust boundary directly affected
by the attack.
& Poisoned information written into long-term agent memory targets
memory or long-term state. \\

\rowcolor{gray!10}
& Attacker access
& Records the access or control required to introduce the malicious
condition.
& An attacker controls a webpage that is later visited by the agent. \\

\rowcolor{gray!10}
& Delivery channel
& Records the medium through which malicious content or behavior reaches
the agentic system.
& A malicious instruction reaches the agent through MCP metadata or a
tool description. \\

\rowcolor{gray!10}
& Insertion point
& Identifies where malicious content or state is introduced into the
workflow.
& A poisoned memory item is inserted through the memory-write path. \\

\rowcolor{gray!10}
& Activation condition
& Records the event or condition under which the attack becomes
effective.
& A poisoned memory item becomes active only when recalled during a
later task. \\

\addlinespace[4pt]

\textbf{\textsc{Agent behavior}}
& Abused capability
& Identifies the agentic capability whose use makes the attack
security-relevant.
& The agent selects and invokes a malicious tool based on its metadata. \\

& Attacker objective
& Captures the security-relevant outcome sought by the attacker.
& A malicious tool causes the agent to perform an unauthorized action. \\

& Downstream impact
& Captures the resulting security-relevant consequence.
& The agent transmits a credential or private file to an external
endpoint. \\

& Persistence / propagation
& Records whether malicious influence persists beyond the initial
interaction or spreads across agents or components.
& A compromised agent forwards malicious instructions to another agent. \\

\addlinespace[4pt]

\rowcolor{gray!10}
\textbf{\textsc{Evidence}}
& Context
& Records the empirical setting in which the attack or safeguard is
supported.
& An attack demonstrated in an isolated experimental agent environment
is coded as sandbox evidence. \\

\rowcolor{gray!10}
& Evidence level
& Captures the strength of source-supported evidence for the coded
observation.
& A source provides an end-to-end demonstration of successful attack
execution under a controlled configuration. \\

\rowcolor{gray!10}
& Evidence type
& Records the form of evidence supporting the coded observation.
& The source provides a controlled experiment, benchmark result, or
end-to-end proof of concept. \\

\addlinespace[4pt]

\textbf{\textsc{Safeguard}}
& Control type
& Classifies the security function performed by an atomic safeguard.
& Restricting filesystem and network permissions is coded as
containment. \\

& Status
& Records whether a safeguard is proposed, evaluated, or both.
& A mitigation described and experimentally tested by a source is coded
as proposed and evaluated. \\

& Origin
& Identifies whether safeguard evidence comes from its authors,
independent evaluation, or another source.
& A defense proposed in one study and tested independently in another
is coded as independent evaluation. \\

& Enforcement target
& Identifies the component, action, or state directly protected by the
control.
& Validation of MCP tool descriptions before registration targets tool
metadata. \\

\addlinespace[4pt]

\rowcolor{gray!10}
\textbf{\textsc{Coverage}}
& Direct assessment
& Records whether a safeguard was explicitly assessed against a specific
atomic attack.
& An experiment testing tool validation against a tool-poisoning attack
is coded as directly assessed. \\

\rowcolor{gray!10}
& Coverage level
& Records the observed protection of a directly assessed
attack--safeguard relationship.
& A defense that blocks some variants but remains bypassable under other
conditions is coded as partial coverage. \\

\bottomrule
\end{tabularx}
\end{table*}

\section{Worked Example of Coding and Adjudication}
\label{app:adjudication-example}
To illustrate how ambiguous cases were resolved,
we present two documented pre-adjudication
disagreements involving attack classification
and safeguard coding, respectively. These
examples demonstrate the application of the
frozen resolver rules.
\paragraph{Case}
For attack record F006-A03, the reviewed source describes a
trigger-conditioned backdoor operating during the tool-use stage of an
LLM-based agent. The source organizes backdoor behavior across planning,
memory, and tool-use stages; tool use is the stage at which the implanted
backdoor becomes effective.

\paragraph{Independent Coding}
Coder 1 assigned the record to C2 (Tool, MCP, and Action-Layer
Attacks), whereas Coder 2 assigned it to C5 (Backdoors and
Trigger-Conditioned Compromise). The disagreement therefore concerned
the primary causal class rather than only an auxiliary coding dimension.

\paragraph{Resolver Decision}
The resolver retained C5. Under the frozen decision rule, the primary
class is determined by the causal attack mechanism rather than by the
component or stage at which the malicious effect is realized. In this
case, tool use is an activation surface for a previously implanted
trigger-conditioned backdoor; it is not itself a tool-poisoning or
tool-selection mechanism. The backdoor trigger therefore determines
the primary class.
Table~\ref{tab:adjudication_example} summarizes the independent
coder assignments and the resolver's final decision for this case.

\begin{table}[H]
\centering
\caption{\textsc{Worked adjudication example for attack record F006-A03.}}
\label{tab:adjudication_example}
\small
\setlength{\tabcolsep}{4pt}
\renewcommand{\arraystretch}{1.06}

\begin{tabularx}{\columnwidth}{
    L{1.5cm}
    >{\centering\arraybackslash}p{0.9cm}
    >{\raggedright\arraybackslash}X}
\toprule
\rowcolor{white}
\textbf{Stage} &
\textbf{Code} &
\textbf{Rationale} \\
\midrule

\rowcolor{gray!10}
Coder 1
&
C2
&
Tool use is the observable stage at which malicious behavior is
realized. \\

\rowcolor{white}
Coder 2
&
C5
&
The behavior is conditioned on a previously implanted backdoor trigger.
\\

\rowcolor{gray!10}
Resolver
&
\textbf{C5}
&
The primary class follows the causal mechanism; tool use is an
activation surface rather than the cause of compromise. \\

\bottomrule
\end{tabularx}
\end{table}

This example illustrates why carrier, victim component, and activation
surface are retained as orthogonal dimensions rather than used to define
the primary attack class. It also demonstrates why primary-class
adjudication follows structural split/merge decisions and precedes
downstream dimension reconciliation.

\noindent\textbf{Safeguard Adjudication Example.}
For safeguard record F001-S01, the source
\emph{BadAgent}~\cite{wang2024badagent}
evaluates clean-data fine-tuning as a defense
against implanted backdoors. Coder 1 assigned
\texttt{PROPOSED\_AND\_EVALUATED} for evaluation
status and
\texttt{SOURCE\_AUTHORS\_SELF\_EVALUATED}
for evidence origin. Coder 2 assigned
\texttt{EVALUATED} and
\texttt{INDEPENDENT\_EVALUATION},
respectively.

The resolver retained Coder 2's assignments
because the source adopts an existing
clean-data fine-tuning defense rather than
introducing it as a new safeguard.
The authors experimentally assess this defense,
but the backdoor remains effective under
the evaluated conditions. This distinction
separates the origin of a safeguard from
the study that evaluates it.

%% file: references.bib
@inproceedings{ladisa2023sok,
  author      = {Piergiorgio Ladisa and Henrik Plate and Matias Martinez and Olivier
                 Barais},
  title       = {{SoK}: Taxonomy of Attacks on Open-Source Software Supply Chains},
  booktitle   = {Proc. of {IEEE} S\&P},
  pages       = {1509--1526},
  year        = {2023},
  doi         = {10.1109/SP46215.2023.10179304}
}

@inproceedings{greshake2023notwhat,
  author      = {Kai Greshake and Sahar Abdelnabi and Shailesh Mishra and Christoph
                 Endres and Thorsten Holz and Mario Fritz},
  title       = {Not What You've Signed Up For: Compromising Real-World
                 {LLM}-Integrated Applications with Indirect Prompt Injection},
  booktitle   = {Proc. of {ACM} {AISec}},
  pages       = {79--90},
  year        = {2023},
  doi         = {10.1145/3605764.3623985}
}

@article{wang2024survey,
  author      = {Lei Wang and Chen Ma and Xueyang Feng and Zeyu Zhang and Hao Yang and
                 Jingsen Zhang and Zhiyuan Chen and Jiakai Tang and Xu Chen and Yankai
                 Lin and Wayne Xin Zhao and Zhewei Wei and Jirong Wen},
  title       = {A Survey on Large Language Model Based Autonomous Agents},
  journal     = {Frontiers of Computer Science},
  volume      = {18},
  number      = {6},
  pages       = {186345},
  year        = {2024},
  doi         = {10.1007/s11704-024-40231-1}
}

@article{liu2023promptinjection,
  author      = {Yi Liu and Gelei Deng and Yuekang Li and Kailong Wang and Zihao Wang
                 and Xiaofeng Wang and Tianwei Zhang and Yepang Liu and Haoyu Wang and
                 Yan Zheng and Yang Liu},
  title       = {Prompt Injection Attack against {LLM}-Integrated Applications},
  journal     = {arXiv:2306.05499},
  year        = {2023}
}

@inproceedings{zhao2024models,
  author      = {Jian Zhao and Shenao Wang and Yanjie Zhao and Xinyi Hou and Kailong
                 Wang and Peiming Gao and Yuanchao Zhang and Chen Wei and Haoyu Wang},
  title       = {Models Are Codes: Towards Measuring Malicious Code Poisoning Attacks
                 on Pre-trained Model Hubs},
  booktitle   = {Proc. of {IEEE}/{ACM} {ASE}},
  pages       = {2087--2098},
  year        = {2024},
  doi         = {10.1145/3691620.3695271}
}

@inproceedings{wang2025model,
  author      = {Hao Wang and Shangwei Guo and Jialing He and Hangcheng Liu and
                 Tianwei Zhang and Tao Xiang},
  title       = {Model Supply Chain Poisoning: Backdooring Pre-trained Models via
                 Embedding Indistinguishability},
  booktitle   = {Proc. of {ACM} Web Conf.},
  pages       = {840--851},
  year        = {2025},
  doi         = {10.1145/3696410.3714624}
}

@inproceedings{zou2025poisonedrag,
  author      = {Wei Zou and Runpeng Geng and Binghui Wang and Jinyuan Jia},
  title       = {{PoisonedRAG}: Knowledge Corruption Attacks to {Retrieval-Augmented}
                 Generation of Large Language Models},
  booktitle   = {Proc. of {USENIX} Security},
  pages       = {3827--3844},
  year        = {2025},
  url         = {https://www.usenix.org/conference/usenixsecurity25/presentation/zou-poisonedrag}
}

@article{wang2026mcptox,
  author      = {Zhiqiang Wang and Yichao Gao and Yanting Wang and Suyuan Liu and
                 Haifeng Sun and Haoran Cheng and Guanquan Shi and Haohua Du and
                 Xiangyang Li},
  title       = {{MCPTox}: A Benchmark for Tool Poisoning on Real-World {MCP} Servers},
  journal     = {Proc. of {AAAI}},
  volume      = {40},
  number      = {42},
  pages       = {35811--35819},
  year        = {2026},
  doi         = {10.1609/aaai.v40i42.40895}
}

@article{lee2024promptinfection,
  author      = {Donghyun Lee and Mo Tiwari},
  title       = {Prompt Infection: {LLM}-to-{LLM} Prompt Injection within Multi-Agent
                 Systems},
  journal     = {arXiv:2410.07283},
  year        = {2024}
}

@inproceedings{arif2026conjunctive,
  author      = {Nokimul Hasan Arif and Qian Lou and Mengxin Zheng},
  title       = {Conjunctive Prompt Attacks in Multi-Agent {LLM} Systems},
  booktitle   = {Proc. of {ACL}},
  pages       = {34175--34191},
  year        = {2026},
  doi         = {10.18653/v1/2026.acl-long.1577}
}

@article{cina2022wild,
  author      = {Antonio Emanuele Cin{\`a} and Kathrin Grosse and Ambra Demontis and
                 Sebastiano Vascon and Werner Zellinger and Bernhard A. Moser and
                 Alina Oprea and Battista Biggio and Marcello Pelillo and Fabio Roli},
  title       = {Wild Patterns Reloaded: A Survey of Machine Learning Security against
                 Training Data Poisoning},
  journal     = {ACM Computing Surveys},
  volume      = {55},
  number      = {13s},
  pages       = {1--39},
  year        = {2023},
  doi         = {10.1145/3585385}
}

@inproceedings{liu2026maliciousskills,
  author    = {Yi Liu and Zhihao Chen and Yanjun Zhang
               and Gelei Deng and Yuekang Li
               and Jianting Ning and Leo Yu Zhang},
  title     = {Do Not Mention This to the User:
               Detecting and Understanding Malicious Agent Skills in the Wild},
  booktitle = {35th USENIX Security Symposium
               (USENIX Security 26)},
  year      = {2026},
  pages     = {1727--1746},
  publisher = {USENIX Association},
  url       = {https://www.usenix.org/conference/usenixsecurity26/presentation/liu-yi}
}

@article{schmotz2026skillinject,
  author      = {David Schmotz and Luca Beurer-Kellner and Sahar Abdelnabi and Maksym
                 Andriushchenko},
  title       = {{Skill-Inject}: Measuring Agent Vulnerability to Skill File Attacks},
  journal     = {arXiv:2602.20156},
  year        = {2026}
}

@techreport{bennet2024aibom,
  author      = {Karen Bennet and Gopi Krishnan Rajbahadur and Arthit Suriyawongkul
                 and Kate Stewart},
  title       = {Implementing {AI} Bill of Materials ({AI BOM}) with {SPDX} 3.0: A
                 Comprehensive Guide to Creating {AI} and Dataset Bill of Materials},
  institution = {The Linux Foundation},
  year        = {2024},
  doi         = {10.70828/RNED4427}
}

@inproceedings{mitchell2019modelcards,
  author      = {Margaret Mitchell and Simone Wu and Andrew Zaldivar and Parker Barnes
                 and Lucy Vasserman and Ben Hutchinson and Elena Spitzer and Inioluwa
                 Deborah Raji and Timnit Gebru},
  title       = {Model Cards for Model Reporting},
  booktitle   = {Proc. of {FAccT}},
  pages       = {220--229},
  year        = {2019},
  doi         = {10.1145/3287560.3287596}
}

@article{gebru2021datasheets,
  author      = {Timnit Gebru and Jamie Morgenstern and Briana Vecchione and Jennifer
                 Wortman Vaughan and Hanna Wallach and Hal {Daum{\'e} III} and Kate
                 Crawford},
  title       = {Datasheets for Datasets},
  journal     = {Communications of the ACM},
  volume      = {64},
  number      = {12},
  pages       = {86--92},
  year        = {2021},
  doi         = {10.1145/3458723}
}

@inproceedings{wang2024badagent,
  author      = {Yifei Wang and Dizhan Xue and Shengjie Zhang and Shengsheng Qian},
  title       = {{BadAgent}: Inserting and Activating Backdoor Attacks in {LLM} Agents},
  booktitle   = {Proc. of {ACL}},
  pages       = {9811--9827},
  year        = {2024},
  doi         = {10.18653/v1/2024.acl-long.530}
}

@inproceedings{yang2024watch,
  author      = {Wenkai Yang and Xiaohan Bi and Yankai Lin and Sishuo Chen and Jie
                 Zhou and Xu Sun},
  title       = {Watch Out for Your Agents! Investigating Backdoor Threats to
                 {LLM}-Based Agents},
  booktitle   = {Proc. of {NeurIPS}},
  volume      = {37},
  pages       = {100938--100964},
  year        = {2024},
  doi         = {10.52202/079017-3201}
}

@article{ji2026cloak,
  author      = {Zimo Ji and Congying Xu and Zongjie Li and Yudong Gao and Xin Wei and
                 Shuai Wang and Shing-Chi Cheung},
  title       = {Cloak and Detonate: Scanner Evasion and Dynamic Detection of Agent
                 Skill Malware},
  journal     = {arXiv:2607.02357},
  year        = {2026}
}

@article{isbarov2026agentproxy,
  author      = {Jafar Isbarov and Murat Kantarcioglu},
  title       = {Bypassing {AI} Control Protocols via Agent-as-a-Proxy Attacks},
  journal     = {arXiv:2602.05066},
  year        = {2026}
}

@article{krishna2025phantoms,
  author      = {Arjun Krishna and Erick Galinkin and Leon Derczynski and Jeffrey
                 Martin},
  title       = {Importing Phantoms: Measuring {LLM} Package Hallucination
                 Vulnerabilities},
  journal     = {arXiv:2501.19012},
  year        = {2025}
}

@inproceedings{chen2024agentpoison,
  author      = {Zhaorun Chen and Zhen Xiang and Chaowei Xiao and Dawn Song and Bo Li},
  title       = {{AgentPoison}: Red-teaming {LLM} Agents via Poisoning Memory or
                 Knowledge Bases},
  booktitle   = {Proc. of {NeurIPS}},
  volume      = {37},
  pages       = {130185--130213},
  year        = {2024},
  doi         = {10.52202/079017-4136}
}

@article{kim2026attackdefenselandscape,
  author      = {Juhee Kim and Xiaoyuan Liu and Zhun Wang and Shi Qiu and Bo Li and
                 Wenbo Guo and Dawn Song},
  title       = {The Attack and Defense Landscape of Agentic {AI}: A Comprehensive
                 Survey},
  journal     = {arXiv:2603.11088},
  year        = {2026}
}

@article{chhabra2026agenticsecurity,
  author      = {Anshuman Chhabra and Shrestha Datta and Shahriar Kabir Nahin and
                 Prasant Mohapatra},
  title       = {Agentic {AI} Security: Threats, Defenses, Evaluation, and Open
                 Challenges},
  journal     = {IEEE Access},
  volume      = {14},
  pages       = {49455--49482},
  year        = {2026},
  doi         = {10.1109/ACCESS.2026.3675554}
}

@article{dehghantanha2026attacksurface,
  author      = {Ali Dehghantanha and Sajad Homayoun},
  title       = {{SoK}: The Attack Surface of Agentic {AI}---Tools, and Autonomy},
  journal     = {arXiv:2603.22928},
  year        = {2026}
}

@article{Zou2023UniversalAdversarial,
  author      = {Andy Zou and Zifan Wang and Nicholas Carlini and Milad Nasr and J.
                 Zico Kolter and Matt Fredrikson},
  title       = {Universal and Transferable Adversarial Attacks on Aligned Language
                 Models},
  journal     = {arXiv:2307.15043},
  year        = {2023}
}

@inproceedings{Kurita2020TrojanLLMs,
  author      = {Keita Kurita and Paul Michel and Graham Neubig},
  title       = {Weight Poisoning Attacks on Pretrained Models},
  booktitle   = {Proc. of {ACL}},
  pages       = {2793--2806},
  year        = {2020},
  doi         = {10.18653/v1/2020.acl-main.249}
}

@inproceedings{Carlini2023PoisoningLLMs,
  author      = {Alvin Chan and Yi Tay and Yew-Soon Ong and Aston Zhang},
  title       = {Poison Attacks against Text Datasets with Conditional Adversarially
                 Regularized Autoencoder},
  booktitle   = {Findings of {EMNLP}},
  pages       = {4175--4189},
  year        = {2020},
  doi         = {10.18653/v1/2020.findings-emnlp.373}
}

@article{Xu2024BackdoorLLM,
  author      = {Yige Li and Hanxun Huang and Yunhan Zhao and Xingjun Ma and Jun Sun},
  title       = {{BackdoorLLM}: A Comprehensive Benchmark for Backdoor Attacks and
                 Defenses on Large Language Models},
  journal     = {arXiv:2408.12798},
  year        = {2024}
}

@article{Sneh2025ToolTweak,
  author      = {Jonathan Sneh and Ruomei Yan and Jialin Yu and Philip Torr and Yarin
                 Gal and Sunando Sengupta and Eric Sommerlade and Alasdair Paren and
                 Adel Bibi},
  title       = {{ToolTweak}: An Attack on Tool Selection in {LLM}-based Agents},
  journal     = {arXiv:2510.02554},
  year        = {2025}
}

@misc{OWASP2025SupplyChain,
  author      = {{OWASP Foundation}},
  title       = {{LLM03:2025 Supply Chain}},
  howpublished= {OWASP},
  year        = {2025},
  url         = {https://genai.owasp.org/llmrisk/llm032025-supply-chain/},
  note        = {Accessed Jul. 20, 2026}
}

@inproceedings{Zhang2025AgentSecurityBench,
  author      = {Hanrong Zhang and Jingyuan Huang and Kai Mei and Yifei Yao and
                 Zhenting Wang and Chenlu Zhan and Hongwei Wang and Yongfeng Zhang},
  title       = {Agent Security Bench ({ASB}): Formalizing and Benchmarking Attacks
                 and Defenses in {LLM}-based Agents},
  booktitle   = {Proc. of {ICLR}},
  year        = {2025},
  url         = {https://openreview.net/forum?id=V4y0CpX4hK}
}

@article{He2024SecurityAIAgents,
  author      = {Yifeng He and Ethan Wang and Yuyang Rong and Zifei Cheng and Hao Chen},
  title       = {Security of {AI} Agents},
  journal     = {arXiv:2406.08689},
  year        = {2024}
}

@article{Deng2026LLMAgentSecurity,
  author      = {Kexin Chu},
  title       = {A Systematic Survey of Security Threats and Defenses in {LLM}-Based
                 {AI} Agents: A Layered Attack Surface Framework},
  journal     = {arXiv:2604.23338},
  year        = {2026}
}

@misc{OWASP2025PromptInjection,
  author      = {{OWASP Foundation}},
  title       = {{LLM01:2025 Prompt Injection}},
  howpublished= {OWASP},
  year        = {2025},
  url         = {https://genai.owasp.org/llmrisk/llm01-prompt-injection/},
  note        = {Accessed Jul. 20, 2026}
}

@inproceedings{Debenedetti2024AgentDojo,
  author      = {Edoardo Debenedetti and Jie Zhang and Mislav Balunovi{\'c} and Luca
                 Beurer-Kellner and Marc Fischer and Florian Tram{\`e}r},
  title       = {{AgentDojo}: A Dynamic Environment to Evaluate Prompt Injection
                 Attacks and Defenses for {LLM} Agents},
  booktitle   = {Proc. of {NeurIPS} Datasets and Benchmarks},
  year        = {2024},
  url         = {https://openreview.net/forum?id=m1YYAQjO3w}
}

@inproceedings{Zhan2024InjecAgent,
  author      = {Qiusi Zhan and Zhixiang Liang and Zifan Ying and Daniel Kang},
  title       = {{InjecAgent}: Benchmarking Indirect Prompt Injections in
                 Tool-Integrated Large Language Model Agents},
  booktitle   = {Findings of {ACL}},
  pages       = {10471--10506},
  year        = {2024},
  doi         = {10.18653/v1/2024.findings-acl.624}
}

@inproceedings{Xie2025ToolSafety,
  author      = {Yuejin Xie and Youliang Yuan and Wenxuan Wang and Fan Mo and Jianmin
                 Guo and Pinjia He},
  title       = {{ToolSafety}: A Comprehensive Dataset for Enhancing Safety in
                 {LLM}-Based Agent Tool Invocations},
  booktitle   = {Proc. of {EMNLP}},
  pages       = {14135--14156},
  year        = {2025},
  doi         = {10.18653/v1/2025.emnlp-main.714}
}

@article{Srivastava2025MemoryGraft,
  author      = {Saksham Sahai Srivastava and Haoyu He},
  title       = {{MemoryGraft}: Persistent Compromise of {LLM} Agents via Poisoned
                 Experience Retrieval},
  journal     = {arXiv:2512.16962},
  year        = {2025}
}

@inproceedings{Shin2025WorkflowProvenance,
  author      = {Souza, Renan and Poteet, Timothy and Etz, Brian and
                 Rosendo, Daniel and Gueroudji, Amal and Shin, Woong and
                 Balaprakash, Prasanna and Ferreira da Silva, Rafael},
  title       = {{LLM} Agents for Interactive Workflow Provenance: Reference
                 Architecture and Evaluation Methodology},
  booktitle   = {Proc. of {SC} Workshops},
  pages       = {2257--2268},
  year        = {2025},
  doi         = {10.1145/3731599.3767582}
}

@article{Ouyang2026MemLineage,
  author      = {Ciyan Ouyang and Rui Hou},
  title       = {{MemLineage}: Lineage-Guided Enforcement for {LLM} Agent Memory},
  journal     = {arXiv:2605.14421},
  year        = {2026}
}

@article{Leong2026ForensicTrajectory,
  author      = {Jun Wen Leong},
  title       = {Forensic Trajectory Signatures for Agent Memory Poisoning Detection},
  journal     = {arXiv:2606.30566},
  year        = {2026}
}

@inproceedings{Rahman2025TestBeforeDeploy,
  author    = {Mohd Sameen Chishti and
               Damilare Peter Oyinloye and
               Jingyue Li},
  title     = {Test Before You Deploy:
               Governing Updates in the {LLM} Supply Chain},
  booktitle = {Companion Proceedings of the 34th ACM
               International Conference on the
               Foundations of Software Engineering
               (FSE Companion '26)},
  year      = {2026},
  pages     = {1584--1587},
  publisher = {ACM},
  doi       = {10.1145/3803437.3805535},
  url       = {https://doi.org/10.1145/3803437.3805535}
}

@article{Jamshidi2025MCP-Guard,
  author      = {Saeid Jamshidi and Arghavan Moradi Dakhel and Kawser Wazed Nafi and
                 Foutse Khomh},
  title       = {Semantic Attacks on Tool-Augmented {LLM}s: Securing the Model Context
                 Protocol Against Descriptor-Level Manipulation},
  journal     = {arXiv:2512.06556},
  year        = {2025}
}

@article{Latendresse2026SkillDependency,
  author      = {Changguo Jia and Tianqi Zhao and Runzhi He and Minghui Zhou},
  title       = {Skills Are Not Islands: Measuring Dependency and Risk in Agent Skill
                 Supply Chains},
  journal     = {arXiv:2607.01136},
  year        = {2026}
}

@article{Agentic2025DefenseSupplyChain,
  author      = {Toqeer Ali Syed and Mohammad Riyaz Belgaum and Salman Jan and
                 Asadullah Abdullah Khan and Saad Said Alqahtani},
  title       = {Agentic {AI} for Autonomous Defense in Software Supply Chain
                 Security: Beyond Provenance to Vulnerability Mitigation},
  journal     = {arXiv:2512.23480},
  year        = {2025}
}

@article{Zhang2025MemorySurvey,
  author      = {Zehao Lin and Xixuan Hao and Renyu Fu and Shaobo Cui and Kai Chen and
                 Chunyu Li and Zhiyu Li and Feiyu Xiong},
  title       = {A Survey on Long-Term Memory Security in {LLM} Agents: Attacks,
                 Defenses, and Governance Across the Memory Lifecycle},
  journal     = {arXiv:2604.16548},
  year        = {2026}
}

@article{Xiong2026CrossSessionInjection,
  author      = {Yuanbo Xie and Tianyun Liu and Yingjie Zhang and Suchen Liu and Yulin
                 Li and Liya Su and Tingwen Liu},
  title       = {What If Prompt Injection Never Left? Exploring Cross-Session Stored
                 Prompt Injection in Agentic Systems},
  journal     = {arXiv:2606.04425},
  year        = {2026}
}

@misc{Koi2026ClawHavoc,
  author      = {Alex and Oren Yomtov},
  title       = {{ClawHavoc}: 341 Malicious Clawed Skills Found by the Bot They Were
                 Targeting},
  howpublished= {Koi Research},
  year        = {2026},
  url         = {https://www.koi.ai/blog/clawhavoc-341-malicious-clawedbot-skills-found-by-the-bot-they-were-targeting},
  note        = {Accessed Jul. 20, 2026}
}

@inproceedings{jiang2026soktaxonomyattackvectors,
  author    = {Xiaochong Jiang and Shiqi Yang and Wenting Yang
               and Yichen Liu and Cheng Ji},
  title     = {Agentic {AI} as a Cybersecurity Attack Surface:
               Threats, Exploits, and Defenses in Runtime Supply Chains},
  booktitle = {2026 IEEE Conference on Artificial Intelligence (CAI)},
  year      = {2026},
  pages     = {2142--2149},
  doi       = {10.1109/CAI68641.2026.11536564}
}

@article{liu2026whenmanuallies,
  author = {Shi Liu and Xuehai Tang and Xikang Yang and
            Liang Lin and Biyu Zhou and Wenjie Xiao and
            Wantao Liu},
  title = {When the Manual Lies: A Realistic Benchmark
           to Evaluate {MCP} Poisoning Attacks for
           {LLM} Agents},
  journal = {arXiv preprint arXiv:2605.24069},
  year = {2026},
  url = {https://arxiv.org/abs/2605.24069}
}
